\documentclass[article]{aa}

\usepackage{graphicx}
\usepackage{txfonts}
\usepackage{hyperref}
\usepackage{upgreek}
\usepackage{color}

\let\originaleqref\eqref
\renewcommand{\eqref}{Eq.~\originaleqref}

\newcommand{\sectionname}{Sect.}

\newcommand{\derivp} [2] {\frac {\partial #1 } {\partial #2} }
\newcommand{\deriv} [2] {\frac {\textrm{d} #1 } {\textrm{d} #2} }

\newcommand{\eq}[1] {Eq.\,(\ref{#1})}

\begin{document}

   \title{
Can plume-induced internal gravity waves regulate the core rotation of subgiant stars?}
  \titlerunning{Angular momentum transport by plume-induced IGW}

   \author{C. Pin\c con\inst{1}, K. Belkacem\inst{1}, M. J. Goupil\inst{1} \and J. P. Marques\inst{2} 
          }

   \institute{LESIA, Observatoire de Paris, PSL Research University, CNRS, Universit\'e Pierre et Marie Curie,
   	Universit\'e Paris Diderot,  92195 Meudon, France\\
	\email{charly.pincon@obspm.fr}
	\and
	Institut d’Astrophysique Spatiale, UMR8617, CNRS, Universit\'e Paris XI, B\^atiment 121, 91405 Orsay Cedex, France
          }

   \mail{charly.pincon@obspm.fr}
   \date{\today}
      
  \authorrunning{C. Pin\c con et al.}

 
  \abstract
   {The seismic data provided by the space-borne missions CoRoT and {\it Kepler} enabled us to probe the internal rotation of thousands of evolved low-mass stars. Subsequently, several studies showed that current stellar evolution codes are unable to reproduce the low core rotation rates observed in these stars. These results indicate that an additional angular momentum transport process is necessary to counteract the spin up due to the core contraction during the post-main sequence evolution. For several candidates, the transport induced by internal gravity waves (IGW) could play a non-negligible role.}
   {We aim to investigate the effect of IGW generated by penetrative convection on the internal rotation of low-mass stars from the subgiant branch to the beginning of the red giant branch. }
   {A semi-analytical excitation model was used to estimate the angular momentum wave flux. The characteristic timescale associated with the angular momentum transport by IGW was computed and compared to the contraction timescale throughout the radiative region of stellar models at different evolutionary stages.}
   {We show that IGW can efficiently counteract the contraction-driven spin up of the core of subgiant stars if the amplitude of the radial-differential rotation (between the center of the star and the top of the radiative zone) is higher than a threshold value. This threshold depends on the evolutionary stage and is comparable to the differential rotation rates inferred for a sample of subgiant stars observed by the satellite {\it Kepler}. Such an agreement can therefore be interpreted as the consequence of a regulation mechanism driven by IGW. This result is obtained under the assumption of a smooth rotation profile in the radiative region and holds true even if a wide range of values is considered for the parameters of the generation model. In contrast, on the red giant branch, we find that IGW remain insufficient, on their own, to explain the observations because of an excessive radiative damping.}
   {IGW generated by penetrative convection are able to efficiently extract angular momentum from the core of stars on the subgiant branch and accordingly have to be taken into account. Moreover, agreements with the observations reinforce the idea that their effect is essential to regulate the amplitude of the radial-differential rotation in subgiant stars. On the red giant branch, another transport mechanism must likely be invoked.}


  \maketitle

%
\section{Introduction}
%

The detection of mixed modes in thousands of stars \citep[e.g.,][]{Mosser2012a} observed by the satellites CoRoT and {\it Kepler} have provided insights into the internal rotation of evolved stars.
Ensemble asteroseismology notably revealed that the mean core rotation weakly increases with time on the subgiant branch \citep{Deheuvels2012,Deheuvels2014} before dropping sharply along the evolution on the red giant branch \citep{Mosser2012b}. These observations have had a major impact since they question our knowledge about the angular momentum evolution and redistribution in stellar interiors. Indeed, the current stellar evolution codes taking transport of angular momentum by shear-induced turbulence and meridional circulation into account are not able to reproduce the low rotation rates observed in evolved stars \citep[][]{Eggenberger2012,Ceillier2013,Marques2013}. These processes are also insufficient to explain the quasi solid-body rotation observed in the solar radiative zone \citep[e.g.,][]{Brown1989,Garcia2007}. These discrepancies
stress out the need for considering additional transport mechanisms to efficiently extract angular momentum from the stellar cores. Among the hypotheses advanced to explain the spin down of the core, several works investigated the effect of magnetic fields \citep[e.g.,][]{Spruit2002,Heger2005,Cantiello2014,Rudiger2015,Spada2016} or the influence of mixed modes \citep{Belkacem2015b,Belkacem2015a}. In addition, it has been known for decades that a possible solution may also come from internal gravity waves \citep[e.g.,][]{Schatzman1996}. 

The restoring force of internal gravity waves (IGW) is buoyancy so that they can propagate in radiative zones of stars. In these regions, they are damped by radiative diffusion and can deposit (or extract) momentum. They are thus able to transport angular momentum in presence of differential rotation. The efficiency of this mechanism depends on the wave energy spectrum whose amplitude and shape results, on the one hand, from the excitation process, and on the other hand, from the wave damping. Two processes have been proposed to generate IGW: turbulent pressure in the convective bulk \citep{Press1981,Garcia1991,Zahn1997,Kumar1999,Lecoanet2013} or penetrative convection at the interface between the convective and the radiative zones \citep{Townsend1966,Pincon2016}. For both mechanisms, excitation models are available to estimate the wave energy flux emitted at the base of the convective zone. These models have already demonstrated the ability of IGW to efficiently affect the evolution of the internal rotation in the solar case~\citep{Zahn1997,Talon2002,Talon2005,Fuller2014,Pincon2016}.~This reinforces the idea that IGW could be one important missing ingredient in the modeling of the angular momentum evolution of stars.

Solving the transport of angular momentum equations including the effects of IGW is numerically difficult. Indeed, it involves very small timescales differing from the stellar evolution timescale by several orders of magnitude. A first step in the investigation thus consists in estimating and comparing the efficiency of the angular momentum transport by IGW with the other transport processes, for a given stellar equilibrium structure and at a peculiar evolutionary stage. Since the core contraction drives the increase of the inner rotation in current stellar modeling, this is equivalent to compare the characteristic timescale associated with the transport by IGW to the contraction timescale. This has already been done in the solar case by previous works \citep{Zahn1997,Kumar1999,Fuller2014,Pincon2016}. Such a comparison provides a first indication on the efficiency of the angular momentum transport by IGW in stars at different evolutionary stages.

\cite{Fuller2014} followed such an approach in the case of evolved low-mass stars. Considering only {\it turbulent pressure in the convective zone} as the excitation mechanism, they showed that the inner radiative interior may decouple from the effect of incoming IGW when stars reach about the middle of the subgiant branch. This mainly results from the wave radiative damping that crucially increases as stars evolve. As a consequence, it prevents IGW from reaching the innermost layers and therefore, from modifying their rotation.

However, as shown in \cite{Pincon2016} in the solar case, the result of such a study depends both on the wave excitation mechanism and on the amplitude of the radial-differential rotation in the radiative zone. Here, we consider the effect of IGW generated by {\it penetrative convection} and investigate whether they can counteract the core contraction and brake the development of a strong differential rotation in subgiant and red giant stars. The influence of both amplitude and shape of the radial-differential rotation on the transport by IGW toward the central regions is considered. 

The article is organized as follows. Section~\ref{timescale and wave flux} introduces the theoretical background. An analytical description of the angular momentum transport by plume-induced IGW is developed. Section~\ref{input physics} briefly presents the input physics and stellar models used in this work. The efficiency of the transport by plume-induced IGW is investigated in \sectionname{}~\ref{IGW transport}. The study is based on timescales comparison for 1 M$_\odot$ stellar models covering the subgiant branch and the beginning of the ascent of the red giant branch. The influence of the uncertainties on the excitation model parameters and of the shape of the rotation profile on the results is widely explored. In \sectionname{}~\ref{regulation}, the effect of stellar mass is analyzed and a comparison with observations is performed. Section~\ref{Discussion} is devoted to discussions.
Conclusions are finally formulated in \sectionname{}~\ref{Conclusion}.

%
\section{Characteristic timescales and wave flux generated by penetrative convection}
%
\label{timescale and wave flux} 

\subsection{Transport timescales}

The contraction of the central layers during the post-main sequence turns out to be mostly responsible for the strong rotation contrast between the core and the envelope observed in stellar evolution codes \citep[e.g.,][]{Ceillier2013,Marques2013}. To estimate the efficiency of the angular momentum transport by IGW in stellar interiors, we can thus compare the influence of IGW versus the contraction or expansion in stars on the evolution of the internal rotation. The angular momentum transport in stellar interiors obeys an advection-diffusion equation. Within the shellular rotation approximation \citep{Zahn1992} and considering only the effect of both the contraction or expansion and the transport by waves, the Lagrangian evolution of the mean rotation rate, $\Omega(r)$, in a spherical mass shell, can be written such as \citep[e.g.,][]{Belkacem2015a}
\begin{align}
\deriv{\Omega}{ {\rm t}} = - 2 \frac{\dot{r}}{r} \Omega - \frac{ \dot{J}_{{\rm w}}}{\rho r^2}\mbox{ ,}
\label{domega_dt}
\end{align}
with
\begin{align}
 \dot{J}_{{\rm w}} = \frac{1}{r^2} \derivp{}{r}\left( r^2 \mathcal{F}_{J,\rm{w}} \right) \mbox{ ,}
\label{J_dot}
\end{align}
where $r$ is the radius, $\dot{r}$ is the contraction or expansion velocity, $\rho$ is the density of the equilibrium structure and $\mathcal{F}_{J,\rm{w}}$ is the mean radial wave flux of angular momentum. 

Following \eqref{domega_dt}, the local contraction or expansion timescale at a radius $r$ is given by
\begin{align}
t_{{\rm cont}} \sim &\left| \frac{r}{2 \dot{r}} \right| \mbox{ .}
\label{t_cont}
\end{align}
It corresponds to the evolution timescale of the rotation rate under the hypothesis of local conservation of angular momentum.
Similarly, the wave-related timescale, or the timescale associated with the transport of angular momentum by IGW, given a rotation profile $\Omega(r)$, can be estimated by
\begin{align}
t_{{\rm w}} \sim\left| \frac{\rho r^2 \Omega}{ \dot{J}_{{\rm w}}}\right| \mbox{ .}
\label{t_w}
\end{align}
It corresponds to the ratio of the density of angular momentum to the divergence of the radial angular momentum wave flux. If $\dot{J}_{{\rm w}}(r)<0$, waves tend to increase the rotation rate following \eqref{domega_dt}, otherwise, they tend to decrease it. The effect of the angular momentum transport induced by IGW on the rotation is said to be locally more efficient than the influence of the contraction or the dilatation in the star if $\epsilon(r)=  t_{{\rm cont}}/t_{{\rm w}}>1$. Therefore, the comparison between \eqref{t_cont} and \eqref{t_w} provides us with an estimate of the efficiency of the angular momentum redistribution by IGW throughout the radiative zone of the stars.

\subsection{Plume-induced wave flux of angular momentum}
\label{plume flux}
The computation of \eqref{t_w} requires the knowledge of the angular momentum wave flux, and hence the wave excitation rate. In this work, we consider only the penetration of convective plumes as the generation mechanism. Convective plumes are strong, coherent, downwards flows that originate from matter cooled at the surface of stars. They grow by turbulent entrainment of matter at their edges when descending through the convective zone \cite[][]{Morton1956}. Once they reach the base of the convective zone, they penetrate by inertia into the underlying stably stratified layers at the top of the radiative region where they are slowed down by buoyancy braking. This is the so-called penetration zone. There, a part of their kinetic energy is converted into waves that can then propagate toward the center of the star \citep[e.g.,][]{Brummell2002, Dintrans2005,Rogers2005,Alvan2015}. To model this process, \cite{Pincon2016} considered the ram pressure exerted by an ensemble of convective plumes in the penetration region as the source term in the wave equation. In the solar case, they found that this mechanism can generate a wave energy flux representing about 1\% of the solar convective flux at the top of the radiative zone. In the following, the mean radial wave energy flux at the top of the radiative zone, per unit of cyclic frequency $\omega$, for an angular degree $l$, and for an azimuthal number $m$, is estimated by Eq.~(39) of their paper,~that is
\begin{align}
\mathcal{F}_{E,\rm{w}}(r_t,\omega,l,m) &\sim  \mathcal{F}_{0}
\frac{e^{-\omega^2/4 \nu_p^2}}{\nu_p} \sqrt{l(l+1)} e^{-l(l+1) b^2/2 r_t^2} \mbox{  ,}
\label{wave energy flux}
\end{align}
with 
\begin{align}
\mathcal{F}_{0} &=-\frac{1}{2\pi}\frac{\mathcal{A}}{4 \pi r_t^2}\frac{ \rho_t V_p^3 \mathcal{S}_p }{2}F_{R} \mbox{  ,}
\label{F_0}
\end{align}
where $r_t$ is the radius at the top of the radiative zone, $\nu_p$ is the plume occurrence frequency,
$b$ is the plume radius, $\mathcal{A}$ is the plumes filling factor, and $\mathcal{S}_p=\pi b^2$ is the horizontal area occupied by one single plume, in the excitation region. $\rho_t$ and $V_p$ are respectively the density and the velocity of the plumes at the base of the convective zone.
The dimensionless quantity $F_R$ is equivalent to the Froude number at the top of the radiative region. It is equal to $F_{R}=\pi V_p/ r_t N_t$, with $N_t$ the Brunt-Väisälä frequency at the top of the radiative zone (see \sectionname{}~\ref{model parameters} for details on its estimate). It controls the efficiency of the energy transfer from the convective plumes into waves inside the penetration region (as it also does in the case of IGW generated by turbulent pressure). By writing \eqref{wave energy flux}, we have assumed that the Péclet number at the base of the convective zone is very high, such that the plumes suffer a strong buoyancy braking leading to a very small penetration length. Indeed, using various stellar models on the subgiant and red giant branches, we have estimated the Péclet number by $P_\mathrm{e} \sim V_p H_t / K_t$, where $V_p$ was computed following \sectionname{}~\ref{model parameters}, $K_t$ is the radiative diffusion coefficient (see below) and $H_t$ the pressure scale height in this region. We have found $P_\mathrm{e}~\sim~10^6-10^7$ for all the models, which justifies the latter assumption \citep[see also the discussion by][]{Dintrans2005}. Moreover, we have assumed a quasi discontinuous profile for the Brunt-Väisälä frequency in this region. Since the smoother the transition of the temperature gradient from an adiabatic to a radiative value at the base of the convective zone, the higher the wave transmission into the radiative zone, this corresponds to consider a lower limit of the wave energy flux regarding the transition length.

In the adiabatic limit, the angular momentum luminosity of each wave component, $\mathcal{L}_{\rm w}$, is conserved when propagating into the stellar interior \citep[e.g.,][]{Bretherton1969,Zahn1997,Ringot1998}, that is
\begin{align}\mathcal{L}_{\rm w}=4\pi r^2 \frac{m}{ \hat{\omega}} \mathcal{F}_{E,\rm{w}}(r,\hat{\omega},l,m) = \rm{const,}\label{lum w}\end{align}
where
\begin{equation}
\hat{\omega}(r,\omega,m)=\omega-m\delta \Omega(r) \label{intrinsic frequency}
\end{equation}
is the Doppler-shifted intrinsic wave frequency\footnote{\label{foot 1}As a convention, we adopt a plane-wave description in the form of $e^{i(\sigma t -m\varphi)}$, with $\sigma$ and $\varphi$ as the wave frequency and the longitudinal coordinate in an inertial frame, respectively. The intrinsic wave frequency in a frame co-rotating with the excitation zone is then given by $\omega = \sigma-m \Omega(r_t)$. Therefore, prograde waves are such as $m\delta\Omega>0$, while retrograde waves are such as \smash{$m\delta\Omega<0$}.} with respect to the excitation site and $\delta \Omega=\Omega(r)-\Omega(r_t)$ is the radial-differential rotation between the radius $r$ and the top of the radiative zone.
Therefore, using \eqref{lum w}, the wave flux of angular momentum can be written in each layer of the radiative zone such as
\begin{align}\frac{m}{ \hat{\omega}} \mathcal{F}_{E,\rm{w}}(r,\hat{\omega},l,m)=\frac{m}{ \omega}  \frac{r_t^2}{r^2} \mathcal{F}_{E,\rm{w}}(r_t,\omega,l,m)\mbox{ .}
\end{align}
Without any dissipative process, we thus recover that $\dot{J}_{{\rm w}}=0$ in \eqref{domega_dt} and that the angular momentum transport by waves is null.

In the non-adiabatic case, each spectral component of the total wave energy flux emitted from the top of the radiative zone undergoes a radiative damping as it propagates toward the center of the star and so contributes to the transport of angular momentum in presence of differential rotation. The wave flux of angular momentum as a function of the depth in the radiative zone can then be derived from the wave energy flux, generated in the excitation region, and modulated by a damping term\footnote{
The right-hand side of Eq. (48) in \cite{Pincon2016} contains a typo and must be corrected by a factor $r_d^2/r^2$ (here, $r_d=r_t$). We note that the numerical computations properly took this factor into account.
} \citep[e.g.,][]{Press1981,Zahn1997} , that is
\begin{align}
\mathcal{F}_{J,\rm{w}}(r)=\sum_l \sum_{m=-l}^{m=+l} \int_{-\infty}^{+\infty}  \frac{m}{\omega} \frac{r_t^2}{r^2} \mathcal{F}_{E,\rm{w}}(r_t,\omega,l,m) e^{-\tau(r,\hat{\omega},l)} \mathrm{d} \omega \mbox{  ,}
\label{total flux}
\end{align}
with
\begin{equation}
\tau(r,\hat{\omega},l)=\left[l(l+1)\right]^{3/2}\int_{r}^{r_t} K  \frac{N N_T^2}{\hat{\omega}^4}\left(\frac{N^2}{N^2-\hat{\omega}^2}\right)^{1/2} \frac{\mathrm{d} r}{r^3} \mbox{  ,}
\label{damping}
\end{equation}
where $N$ is the Brunt-Väisälä frequency, with $N_T$ its thermal part that does not take the gradient in the chemical composition into account, and $K$ (in units of m$^2$~s$^{-1}$) is the radiative diffusion coefficient \citep[e.g][]{Maeder2009}. We note that the wave flux of angular momentum depends on the Doppler-shifted frequency $\hat{\omega}(r)$ only through the wave damping. We also emphasize, following \eqref{damping}, that IGW deposit all the angular momentum they carry just above their critical layers (i.e where $\hat{\omega}=0$) and cannot propagate further downwards.

In the following, we will neglect wave reflection and will assume that each wave component is lost the first time it reaches a reflection point near the center (i.e where $N^2=\hat{\omega}^2$). In other words, it is supposed either to be absorbed into the medium before or to be so weakly damped in its propagation cavity that it does not quantitatively contribute to the transport. Moreover, we stress that the modeling of the transport by IGW represented by Eqs. (\ref{wave energy flux}), (\ref{total flux}) and (\ref{damping}) neglects the Coriolis force and the wave heat flux. The implications of these hypotheses on the results are discussed a posteriori in \sectionname{}~\ref{hypotheses transport}.

%
\section{Input physics}
%
\label{input physics}

\subsection{Stellar models}
\label{stellar models}
\begin{figure}[]
\centering
\includegraphics[scale=0.5,trim= 0.5cm 0cm 0cm 0cm, clip]{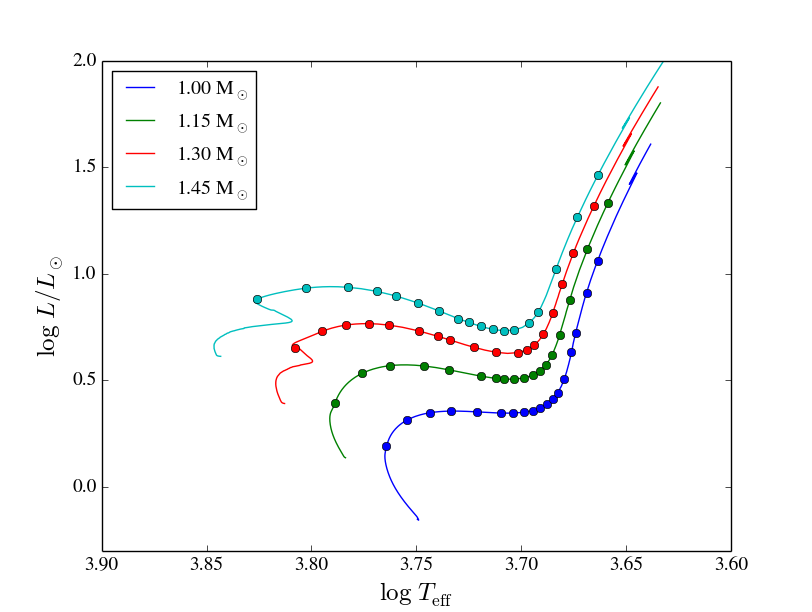}
\caption{Location in the Hertzsprung-Russel diagram of the stellar models considered in this work (filled circles, see \sectionname{}~\ref{stellar models} for details). The evolutionary tracks from the zero-age main sequence to the top of the red giant branch are represented by the solid lines.}
\label{HR}       
\end{figure}
We considered evolutionary sequences of 1, 1.15, 1.3 and 1.45~M$_{\odot}$ models computed with the stellar evolution code CESTAM \citep{Marques2013}. The sequences cover the subgiant branch and the beginning of the ascent of the red giant branch. The mass range was chosen to be representative of observed low-mass subgiant and red giant stars \citep{Mosser2012b,Deheuvels2014}. Their location in the Hertzsprung-Russel diagram is displayed in \figurename{} \ref{HR}. Their chemical composition is similar to the solar mixture as given in \cite{AGS2009}, with initial helium and metals abundances $Y_0=0.25$ and $Z_0=0.013$. We used the OPAL 2005 equation of states and opacity tables. Microscopic diffusion, overshooting and rotation were neglected. The stellar atmosphere was constructed following an Eddington gray approximation and the nuclear reaction rates were deduced from the NACRE compilation, except for the ${}^{14}N(p,\gamma){}^{15}O$ reaction that follows \cite{Imbriani2004}. Finally, convection was modeled using the mixing-length theory \citep{Bohm1958} with $\alpha_{MLT}=1.65$.

\subsection{Excitation model parameters}
\label{model parameters}

The plume radius $b$ and velocity $V_p$ were estimated using the semi-analytical model of turbulent plumes by \cite{Rieutord1995}, as detailed in \cite{Pincon2016}. The value of the Brunt-Väisälä frequency at the top of the radiative zone, $N_t$, remains difficult to estimate in current stellar models. Since $N$ approximately follows a power law with respect to the radius between the hydrogen-burning shell and the outer edge of the radiative zone \citep[e.g.,][]{Mosser2017}, we considered the value given by the extrapolation of this power law at the base of the convective zone (i.e., at $r=r_t$).
The filling factor $\mathcal{A}$ is a proxy for the number of penetrating convective plumes, with an upper limit fixed at 0.5 by mass conservation. Realistic 3D Cartesian numerical simulations of the uppermost convective layers of evolved red giant stars show values around $\mathcal{A}\sim 0.1$ \citep{Ludwig2012}, which is very close to the values obtained for the Sun \citep[e.g.,][]{Stein1998}. This qualitatively agrees with numerical simulations of the extended convective envelope of RGB stars \citep[e.g.,][]{Brun2009,Palacios2012} in which the energy transport by strong cooler downwards convective plumes is well developed and predominant. We then chose $\mathcal{A}\sim 0.1$ as the default value.
Finally, given the lack of knowledge about the plume lifetime, the default value will be taken around the convective turnover timescale as given by the MLT, $\nu_p \sim \omega_c$ (see \figurename{}~\ref{caract}). The effect of variations in the model parameters on the transport by IGW is studied in \sectionname{}~\ref{uncertainties}. We emphasize that \eqref{total flux} was computed in a frequency range between $0$ and $\min(N_t,10~\nu_p)$ and for angular degrees between $l=1$ and $l=150$. These choices were checked a posteriori to be sufficient for all the considered models.

\subsection{Assumptions for the rotation profile}
\label{rotation profile}
\begin{figure}[]
\centering
\includegraphics[scale=0.5,trim= 0.5cm 0cm 0cm 0cm, clip]{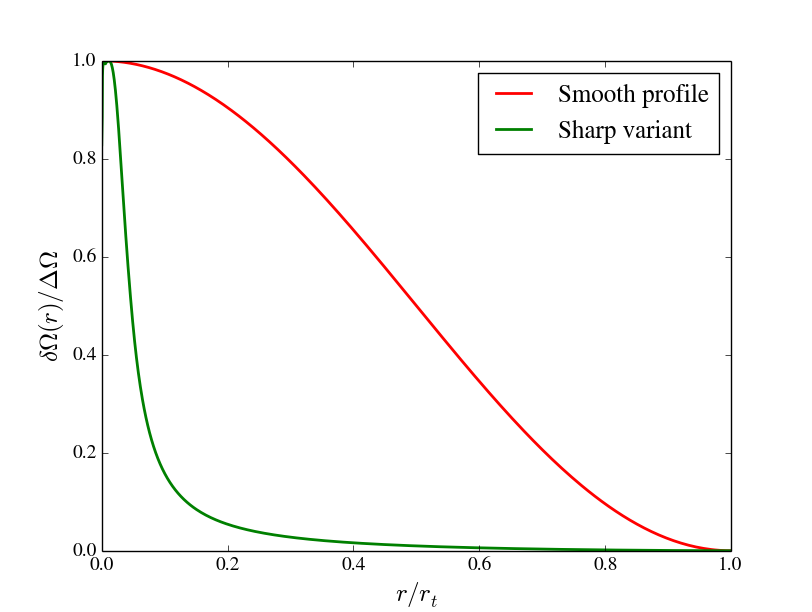}
\caption{Normalized differential rotation profile according to the radius normalized by the radius at the top of the radiative zone, $r_t$, for a 1M$_\odot$ subgiant model with $\log~T_{{\rm eff}}=3.70$ and $\log~L/L_\odot = 0.35$. We compare the smooth profile given in \eqref{smooth domega} (red solid line) with the sharp variant given in \eqref{sharp domega} (green solid line).}
\label{profile_rot}       
\end{figure}
As we can see in Eqs. (\ref{domega_dt}), (\ref{J_dot}), (\ref{total flux}) and (\ref{damping}), the evolution of the internal rotation and the action of IGW are intricately related. In particular, the differential rotation, $\delta \Omega (r)$, plays a major role in determining the wave flux via the radiative damping term. On the one hand, it causes the asymmetry between prograde ($m>0$) and retrograde ($m<0$) waves that is essential for the extraction of angular momentum (indeed, if $\delta \Omega=0$, transport of angular momentum by prograde and retrograde waves cancel each other out so that $\dot{J}_{{\rm w}} =0$). In the other hand, it controls the magnitude of the wave damping through the dependence on $\hat{\omega}^{-4}$ and hence, the local deposit of angular momentum into the medium. As a consequence, it has been shown in the solar case that its amplitude can have a large impact on the transport by IGW \citep{Pincon2016}. Therefore, we will assume a given rotation profile for each stellar model. We thus implicitly consider that the effect of the centrifugal forces on the equilibrium structure is negligible.
While seismic studies can provide stringent constraints on the mean core rotation rate in subgiant and red giant stars, little is known about the shape of the rotation profile in these stars and difficulties remain, in some cases, to discriminate between a sharp rotation profile and a smoother one \cite[][]{Deheuvels2014}. 
We thus consider two limiting cases for the rotation profile: a smooth profile and a sharp one.

\paragraph{Smooth profile:} Seismic observations show evidence that a mechanism is able to prevent the core contraction from developing a large differential rotation in the radiative zone and to efficiently redistribute angular momentum from the core to the envelope all along the star lifetime. As a consequence, it seems reasonable to expect the rotation profile to be smoother than the one obtained by assuming local conservation of angular momentum. A linearly decreasing profile in the radiative zone is the simplest form we could assume. However, to enforce a null derivative at the center and at the top of the radiative zone, we prefer considering a profile with the shape
\begin{align}
\delta \Omega(r)=\Delta \Omega \cos^2\left(\frac{\pi }{2 }\frac{r}{r_t}\right) ~~~~\mathrm{for}~~r\le r_t\mbox{ ,}
\label{smooth domega}
\end{align}
with $\Delta \Omega$ the maximum amplitude of the differential rotation. Equation~(\ref{smooth domega}) will be considered as the default profile.

\paragraph{Sharp profile:} In a second step, we will investigate the effect of a sharp rotation profile. It will be supposed to be similar to the one obtained by assuming local conservation of angular momentum from the beginning of the subgiant branch, but rescaled with a different amplitude. For initial conditions, we will assume that the rotation is uniform throughout the radiative zone at the terminal-age main sequence (TAMS), as supported by observations in main-sequence stars \citep[e.g.,][]{Benomar2015}. The resolution of \eqref{domega_dt} with $\dot{J}_{{\rm w}}=0$ thus gives
\begin{align}
\delta \Omega(r)=\Delta \Omega\frac{ \mathcal{C}^2(m_r)-\mathcal{C}^2(m_{r,t}) }{\max\left[\mathcal{C}^2(m_r)-\mathcal{C}^2(m_{r,t})\right]}~~~~\mathrm{for}~~r<r_t\mbox{ ,}
\label{sharp domega}
\end{align}
with 
\begin{align}
\mathcal{C}(m_r)=\frac{r_0(m_r)}{r(m_r)}\mbox{  ,}
\end{align}
where $r_{0}(m_r)$ is the radius of the mass element $m_r$ at the TAMS and $m_{r,t}$ is the mass coordinate at the top of the radiative zone.
\\

Both kind of rotation profiles are represented in \figurename{} \ref{profile_rot} for a 1M$_\odot$ model in the middle of the subgiant branch. Considering such arbitrary shapes for the differential rotation, although questionable, will give us first hints about the efficiency of the transport by IGW in subgiant and red giant stars while considering different values for the amplitude ${\Delta \Omega}$.
A comparison with the differential rotation amplitude inferred by seismic observations requires accounting for the properties of the gravity-dominated mixed modes. Following \cite{Goupil2013}, the mean seismic amplitude of the differential rotation, that is its amplitude as probed by mixed modes, is given by
\begin{align}
\overline{\Delta \Omega}\approx\frac{1}{\gamma} \int_0^{r_t} \delta \Omega \frac{N}{r} \mathrm{d} r ~~~~\mathrm{with}~~~~\gamma=\int_0^{r_t}  \frac{N}{r} \mathrm{d} r \mbox{ ,}
\label{seismic domega}
\end{align}
which have been derived by assuming that the rotation rate is uniform throughout the convective envelope. Equation~(\ref{seismic domega}) shows that mixed modes probe the mean value of the rotation in a region confined around the peak of the Brunt-Väisälä frequency, as also noted by \cite{Deheuvels2014}. Therefore, a comparison between the values of the differential rotation as observed by mixed modes and the ones as assumed in this work will make sense if we consider the mean seismic amplitude, \smash{$\overline{\Delta  \Omega}$}, computed via \eqref{seismic domega}.

%
\section{Efficiency of the angular momentum transport by IGW in subgiant and red giant stars}
%

\label{IGW transport}

\subsection{Comparison of timescales}
\label{comparison}

\begin{figure}[]
\centering
\includegraphics[scale=0.5,trim= 0.5cm 0cm 0cm 0cm, clip]{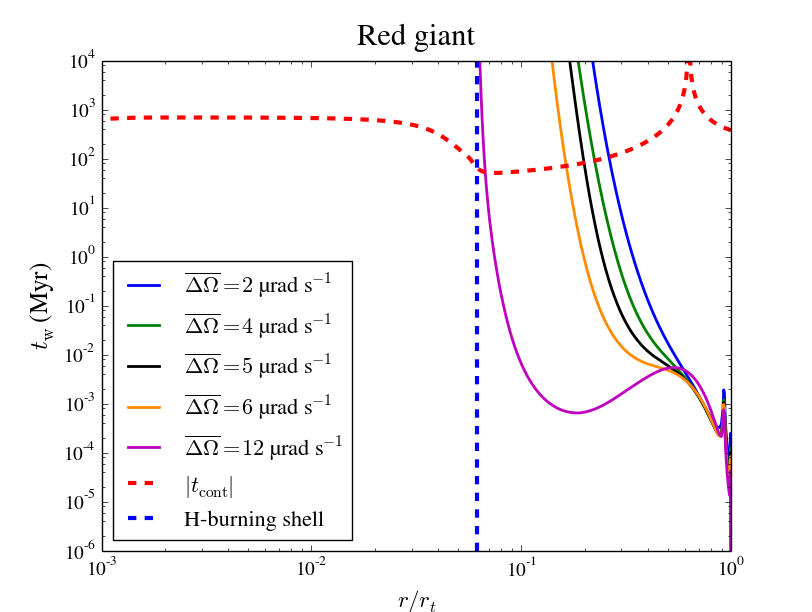}
\includegraphics[scale=0.5,trim= 0.5cm 0cm 0cm 0cm, clip]{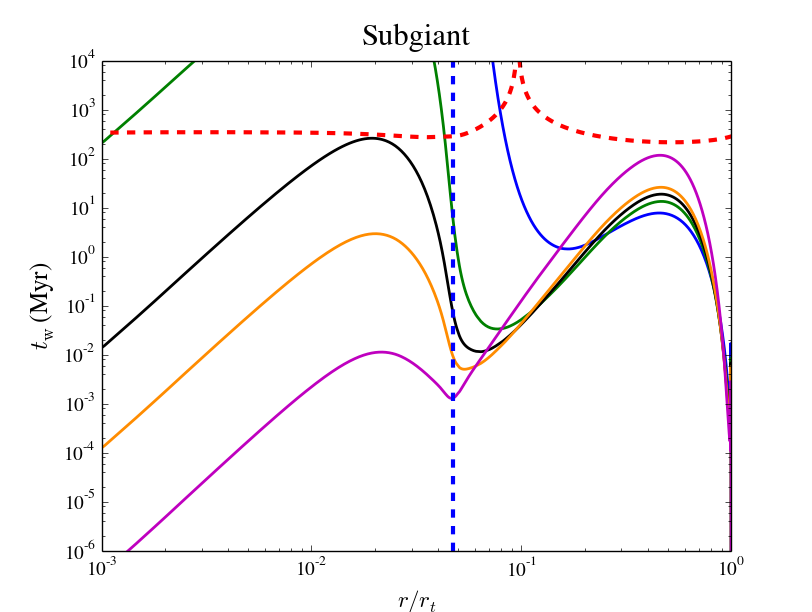}
\caption{{\bf Top:} Characteristic wave-related timescale (solid lines), computed using \eqref{t_w}, as a function of the normalized radius in
the radiative zone of a 1M$_\odot$ red giant model, with $\log~T_{{\rm eff}} =3.16$ and $\log~L/L_\odot = 0.92$. Colors correspond to different values for
the amplitude of the differential rotation, \smash{$\overline{\Delta \Omega}$}. The red and the blue dashed lines represent the contraction or dilatation timescale and the location of the hydrogen-burning shell, respectively. {\bf Bottom:} Same caption and legend as the top panel, but for a 1M$_\odot$ subgiant model, with $\log~T_{{\rm eff}} = 3.70$ and $\log~L/L_\odot = 0.35$.}
\label{timescale sg rg}      
\end{figure}

\subsubsection{Stars at the beginning of the ascent of the RGB}
\label{t_w rg}

As an illustration, we consider the case of a 1M$_\odot$ model characterized by $\log~L/L_\odot =0.92$, $\log~g=3.16$, $\log ~T_{{\rm eff}}=3.67$ and located at the beginning of the ascent of the red giant branch. The wave-related timescale is compared to the contraction timescale throughout the radiative zone of the star in \figurename{}~\ref{timescale sg rg}. The values for \smash{$\overline{\Delta \Omega}$} are chosen between $0$ and $12~\upmu$rad~s$^{-1}$, that is in a range with an upper limit slightly above the maximum value inferred from asteroseismology \citep{Mosser2012b,Deheuvels2014}. 

At the very top of the radiative zone, \figurename{}~\ref{timescale sg rg} shows that $t_{{\rm w}}$ takes very small values of the order of years. This is due to the damping of high-angular degree low-frequency prograde waves ($m>0,\dot{J}_{{\rm w}}<0$) that rapidly reach their critical layers. Just below this region, high-angular degree low-frequency retrograde waves ($m<0, \dot{J}_{{\rm w}}>0$) are in turn efficiently damped. Combined with shear-induced turbulence, this may give rise to a short-period oscillation of the rotation profile in a region confined around the top of the radiative zone, the so-called {\it Shear Layer Oscillation} \citep[e.g.,][]{Kumar1999,Talon2002}. 

Deeper in the star (i.e., $r/r_t\lesssim 0.8$), $t_{{\rm w}}$ drastically increases as $r$ decreases. It even gets much larger than the contraction timescale throughout the helium core, below the hydrogen-burning shell. 
Despite the significant influence of the differential rotation amplitude on $t_{{\rm w}}$ above the helium core, the result remains valid for values of \smash{$\overline{\Delta \Omega}$} consistent with observations.
Unlike main-sequence stars \citep[see][for the solar case]{Pincon2016}, IGW generated by penetrative convection are thus not efficient enough to affect the innermost rotation of red giant stars.
Indeed, as pointed out by \cite{Fuller2014}, the wave radiative damping increases when stars evolve during the post-main sequence. As shown in \appendixname{}~\ref{spatial behavior}, the local characteristic radial damping lengthscale of the wave energy decreases all over the radiative zone as the innermost layers contract. This decrease is so important at the beginning of the RGB that waves are severely damped before reaching the hydrogen-burning shell for values of \smash{$\overline{\Delta \Omega}\lesssim12~\upmu$}rad~s$^{-1}$.
Actually, we find that a value of \smash{$\overline{\Delta \Omega}$} around $20~\upmu$rad~s$^{-1}$ is needed to enable retrograde IGW to reach the helium core in this model, that is a value much higher than observations. We find similar conclusions for all the red giant models considered in \figurename{}~\ref{HR}. Such a trend is further discussed in \sectionname{}~\ref{t_w sg}.
We then conclude that IGW generated by penetrative convection cannot counteract, on their own, the acceleration due to the core contraction in RGB stars and therefore, are unable to spin down their innermost layers. Another mechanism must operate in the core of these stars.

\subsubsection{Subgiant stars}
\label{t_w sg}

As an illustration, we consider the case of 1M$_\odot$ subgiant model characterized by $\log~L/L_\odot =0.35$, $\log~g=3.86$ and $\log ~T_{{\rm eff}}=~3.70$. The wave-related timescale is plotted in \figurename{}~\ref{timescale sg rg} as a function of radius in the radiative zone. For low values of \smash{$\overline{\Delta \Omega}$}, the situation is similar as for RGB stars. IGW are considerably damped and deposit their angular momentum well above the hydrogen-burning shell. However, unlike RGB stars, as the amplitude of the differential rotation increases and becomes larger than a threshold value, denoted \smash{$\overline{\Delta \Omega}_{{\rm th}}$}, $t_{{\rm w}}$ becomes smaller than the contraction timescale throughout the radiative zone. Indeed, as \smash{$\overline{\Delta \Omega}$} increases, the asymmetry between prograde ($m>0$) and retrograde waves ($m<0$) gets more and more pronounced. First, prograde components reach more rapidly their critical layers where they are dissipated. Second, an increase in \smash{$\overline{\Delta \Omega}$} results in an increase in the frequency Doppler-shift of retrograde IGW ($m<0, \dot{J}_{{\rm w}}>0$) so that their damping in \eqref{damping} decreases. They can thus go deeper into the star. In other words, when \smash{$\overline{\Delta \Omega}$} increases, retrograde components with lower frequencies are absorbed into the helium core. Since the excitation wave spectrum decreases as the wave frequency $\omega$ increases, they have higher amplitudes and thus deposit more negative angular momentum. We refer to the toy model presented in \appendixname{}~\ref{model appendix} for a more detailed analysis. In the example plotted in \figurename{}~\ref{timescale sg rg}, \smash{$\overline{\Delta \Omega}_{{\rm th}}\sim5~\upmu$rad s$^{-1}$}. This value is consistent with the mean core rotation rates observed in subgiant stars (see \tablename{}~\ref{table2}). Therefore, we conclude that IGW generated by penetrative convection are able to counteract, on their own, the acceleration due to the core contraction in subgiant stars and therefore, can play a role in the redistribution of angular momentum along the subgiant branch.

Actually, this result can be extended to the whole subgiant branch where the same trend can be observed. The theoretical threshold value, \smash{$\overline{\Delta \Omega}_{{\rm th}}$}, is formally defined as the amplitude of the radial-differential rotation above which IGW can counteract the acceleration due to the core contraction ($t_{{\rm w}} \le t_{{\rm cont}}$) throughout the region below the hydrogen-burning shell. It was determined for the 1M$_\odot$ evolutionary sequence represented in \figurename{}~\ref{HR}. We found that \smash{$\overline{\Delta \Omega}_{{\rm th}}$} increases as the star evolves (see \figurename{}~\ref{threshold var}). Indeed, as shown in \appendixname{}~\ref{spatial behavior}, the intensity of the radiative damping increases by more than two orders of magnitude along the evolution from the subgiant branch to the beginning of the RGB. Considering this effect is the predominant one during the star evolution, it would result in a decrease in the efficiency of the transport by IGW in the innermost layers of the radiative zone if \smash{$\overline{\Delta \Omega}$} remained unchanged (i.e., $t_{{\rm w}} > t_{{\rm cont}}$) since waves would be dissipated into upper layers. Therefore, an increase in the frequency Doppler-shift, and so in \smash{$\overline{\Delta \Omega}_{{\rm th}}$}, is necessary to balance the increase in the magnitude of the radiative damping over time and enable retrograde IGW to reach the helium core and efficiently counteract the acceleration due to its contraction (i.e., $t_{{\rm w}} \le t_{{\rm cont}}$, see also \appendixname{}~\ref{influence of damping} for a more detailed discussion).

\subsection{Impact of the parameter uncertainties and of the shape of the rotation profile}
\label{uncertainties}

\begin{figure}[]
\centering
\includegraphics[scale=0.5,trim= 0.5cm 0cm 0cm 0cm, clip]{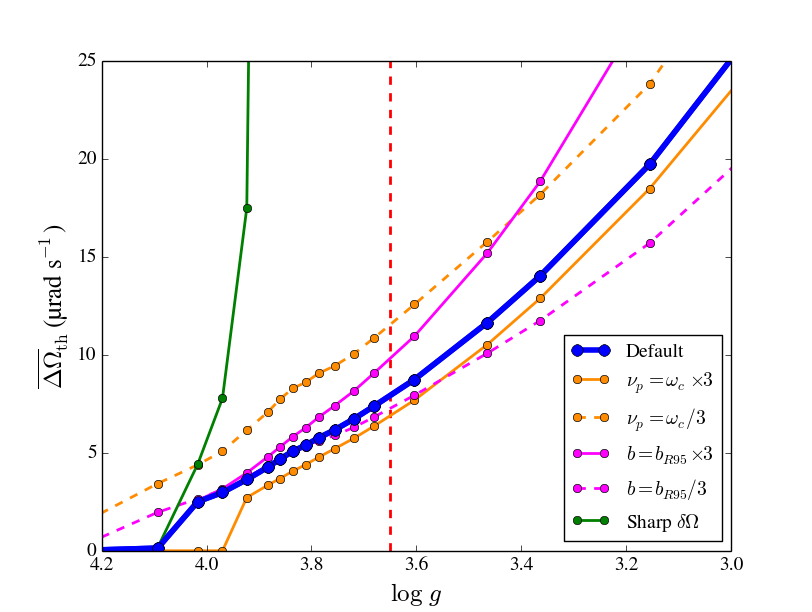}
\caption{Theoretical threshold value for the differential rotation amplitude, above which $t_{{\rm w}} < t_{{\rm cont}}$ throughout the region below the hydrogen-burning shell, as a function of $\log~g$. The curve obtained with the default set of plume parameters, $\nu_p=\omega_c$ and $b=b_{R95}$ (see \sectionname{}~\ref{model parameters}), and a smooth rotation profile is represented by the blue thick solid line. The effect of variations in $\nu_p$ and $b$ is also displayed, as well as the results obtained with a sharp rotation profile. For each curve, the value of only one single parameter is changed while keeping the others at their default values. The vertical red dashed line symbolizes the transition between the subgiant phase to the beginning of the RGB.}
\label{threshold var}     
\centering
\includegraphics[scale=0.5,trim= 0.5cm 0cm 0cm 0cm, clip]{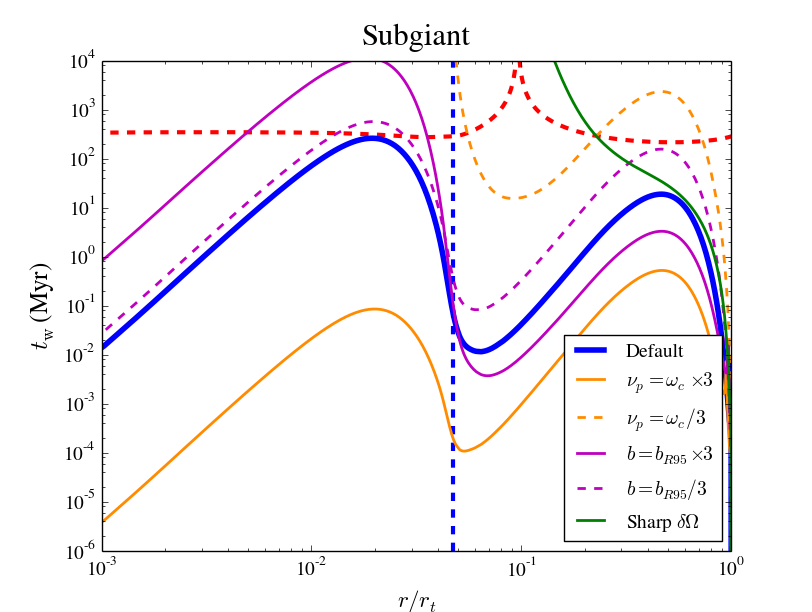}
\caption{Effect of variations in the plume parameters and of a sharp rotation profile as given in \eqref{sharp domega} on the wave-related timescale in the radiative zone of the subgiant model considered in \sectionname{}~\ref{t_w sg}. For each curve, the value of only one single parameter is changed while keeping the others at their default values. The value of the differential rotation amplitude is maintained constant at \smash{$\overline{\Delta \Omega}_{{\rm th}}$}. The red and the blue dashed lines have the same meaning as in \figurename{}~\ref{timescale sg rg}. }
\label{timescale var} 
\end{figure}
Uncertainties remain in the determination of the plume characteristics. First, their estimation relies on simplified analytical models. Second, they have been supposed to be unique for all the plumes at a given evolutionary stage while, in reality, they should be distributed around mean values. In this section, we vary the parameters of the excitation model around the default values used in the previous sections and investigate their influence on the efficiency of the angular momentum transport by IGW. We also explore the effect of a sharp rotation profile on the results.

\paragraph{Wave flux amplitude $\mathcal{F}_0$.}

As shown by Eqs. (\ref{J_dot}), (\ref{t_w}) and (\ref{wave energy flux}), a decrease in $\mathcal{F}_0$ by any factor results in an increase in $t_{{\rm w}}$ by the same factor. This parameter depends on the plume velocity at the base of the convective zone $V_p$, the plume radius via the term $\mathcal{S}_p$, the filling factor (or the number of penetrating plumes), and the efficiency of the energy transfer from the plumes to the waves via the parameter $N_t$. Given that $\mathcal{F}_0$ depends on $V_p$ to the power four, the uncertainty on $V_p$ can result in a large uncertainty in the efficiency of the transport by IGW for a given value of \smash{$\overline{\Delta \Omega}$}. However, as shown in \figurename{}~\ref{timescale sg rg}, $t_{{\rm w}}$ is very sensitive to the amplitude of the differential rotation below the hydrogen-burning shell (see \appendixname{}~\ref{sensitivity} for an explanation). Therefore, a major variation in $\mathcal{F}_0$ leads only to a moderate variation in the threshold \smash{$\overline{\Delta \Omega}_{{\rm th}}$}. For instance, in the subgiant case in \figurename{}~\ref{timescale sg rg}, if $V_p$ is decreased by a factor of three, $t_{{\rm w}}$ increases by about a factor of hundred, but \smash{$\overline{\Delta \Omega}_{{\rm th}}$} increases from $5~\upmu$rad s$^{-1}$ to only about $6~\upmu$rad s$^{-1}$. However, near the top of the radiative zone (i.e., $r/r_t \gtrsim 0.5$), the efficiency of the transport by IGW is more affected. Indeed, an increase in \smash{$\overline{\Delta \Omega}_{{\rm th}}$} leads to an increase in $t_{{\rm w}}$ in this region. Nevertheless, the main conclusions remain unmodified since this region is in expansion ($\dot{r}>0$ in \eqref{domega_dt}) and so spins down. Therefore, the conclusions made in \sectionname{}~\ref{t_w sg} for the subgiant stars hold despite strong variations in $\mathcal{F}_0$. For the red giant case in \figurename{}~\ref{timescale sg rg}, $t_{{\rm w}}$ varies so sharply that the conclusions made in \sectionname{}~\ref{t_w rg} also do not change even if $\mathcal{F}_0$ is multiplied by a factor of hundred. We conclude that, although its effect is quantitatively not negligible, variations in $\mathcal{F}_0$ does not qualitatively modify the results obtained in both subgiant and red giant cases.

\paragraph{Plume width, $b$.}

We modify the plume radius by factors of a third and three. The result is plotted in \figurename{}~\ref{threshold var} and \figurename{}~\ref{timescale var}. As we can see in \figurename{}~\ref{timescale var}, the sensitivity of $t_{{\rm w}}$ to the value of $b$ is moderate throughout the radiative zone. Above the hydrogen-burning shell, $t_{{\rm w}}$ is multiplied by about factors of nine and a ninth, respectively. This trend directly comes from the change in the area occupied by one single plume, $\mathcal{S}_p$ in \eqref{F_0}.
Below the hydrogen-burning shell, the situation is more complex. Indeed, the wave energy flux, \eqref{wave energy flux}, is maximum for an angular degree $l_{{\rm max}} \sim r_t /b$ \citep{Pincon2016}. A smaller value of $b$ then favors the excitation of high-angular degrees and results in a larger width of the wave spectrum with respect to the angular degree. Nevertheless, this must compete with a larger damping at the very top of the radiative zone and a decrease in $\mathcal{S}_p$. Without going more into details, \figurename{}~\ref{threshold var} shows that $b$ has only a small influence on \smash{$\overline{\Delta \Omega}_{{\rm th}}$}.

\paragraph{Plume occurrence frequency, $\nu_p$.}

In a similar way to the plume radius, we modify the plume lifetime by factors of a third and three. Figure~\ref{timescale var} shows that variations in $\nu_p$ have important consequences on the value of $t_{{\rm w}}$. In the subgiant case, increasing (decreasing) $\nu_p$ by a factor of three, while keeping the value of \smash{$\overline{\Delta \Omega}$} constant, leads to decrease (increase) $t_{{\rm w}}$ by more than two orders of magnitude throughout the radiative zone. However, given the abovementioned sensitivity of $t_{{\rm w}}$ to \smash{$\overline{\Delta \Omega}$} below the hydrogen-burning shell, the induced variation in the threshold value \smash{$\overline{\Delta \Omega}_{{\rm th}}$} is much more limited than for $t_{{\rm w}}$. Figure~\ref{threshold var} indicates that it is lower than a factor of two. Hence, the conclusions made in \sectionname{}~\ref{comparison} again qualitatively hold. Figure~\ref{threshold var} also shows that the higher $\nu_p$, the lower \smash{$\overline{\Delta \Omega}_{{\rm th}}$}, whatever the value of $\log$ $g$. Indeed, following \eqref{wave energy flux}, an increase in $\nu_p$ enhances the amplitude of the waves whose frequencies are higher than $\nu_p$ at the expense of the ones whose frequencies are lower than $\nu_p$. The part of the wave spectrum with high enough energy to efficiently modify the innermost rotation is thus extended toward higher frequencies. Therefore, a lower value of the frequency Doppler-shift and so of \smash{$\overline{\Delta \Omega}_{{\rm th}}$} is needed for these waves to counteract the radiative damping, reach the innermost layers and efficiently extract angular momentum from them.

\paragraph{Effect of a sharp rotation profile.}

We also tested the impact of a sharp rotation profile as defined by \eqref{sharp domega}. As shown in \figurename{}~\ref{threshold var} and \figurename{}~\ref{timescale var}, such a profile strongly prevents IGW from reaching the helium core, unless the amplitude of the differential rotation is very large. Compared to a smooth rotation profile (see \figurename{}~\ref{profile_rot}), retrograde IGW are more importantly damped above the hydrogen-burning shell because of a lower frequency Doppler-shift, unlike the prograde ones that can go deeper into the star. At a given value of \smash{$\overline{\Delta \Omega}$}, the asymmetry between prograde and retrograde components is thus reduced compared to the case of a smooth profile. Therefore, once IGW reach the hydrogen burning-shell, the retrograde waves that more efficiently deposit angular momentum into the medium are shifted toward higher frequencies. Since the wave spectrum is a decreasing function of frequency, they deposit much less momentum than in the smooth case. Figure~\ref{threshold var} shows that \smash{$\overline{\Delta \Omega}_{{\rm th}}$} must be in general very large to enable IGW to decelerate the core rotation, with values much higher than the observations on both subgiant and red giant branches. We can see that \smash{$\overline{\Delta \Omega}_{{\rm th}}$} is very small only at the very beginning of the subgiant branch where a strong rotation gradient has not developed yet. Therefore, we conclude that a sharp rotation profile, with a large gradient in the vicinity of the hydrogen-burning shell, prevents IGW from reaching the innermost layers and to modify the inner rotation rate as soon as the beginning of the post-main sequence (i.e., in both subgiant and red giant stars). 
\\

In summary, variations in the plume parameters $\mathcal{F}_0$, $b$ and especially $\nu_p$ have quantitative impacts on the transport by IGW throughout the radiative zone of stars. Nevertheless, they do not qualitatively modify the conclusions made in \sectionname{}~\ref{comparison} for subgiant stars, as well as for red giant stars. Indeed, given the strong sensitivity of $t_{{\rm w}}$ on \smash{$\overline{\Delta \Omega}$} below the hydrogen-burning shell (see \appendixname{}~\ref{sensitivity} for an explanation), large variations in the parameters have only moderate effects on \smash{$\overline{\Delta \Omega}_{{\rm th}}$} that remains within an order of magnitude from the observations in subgiants. We have also demonstrated that a sharp rotation profile, with a large rotation gradient near the hydrogen-burning shell, prevents IGW from spinning down, on their own, the helium core in the post-main sequence.

\section{Observational evidence for a wave-driven regulation mechanism in subgiant stars?}
\label{regulation}

\subsection{Influence of the stellar mass}
\label{influence mass}
We also investigated the influence of the stellar mass on the results provided in \sectionname{}~\ref{IGW transport}. To do so, we considered the 1, 1.15, 1.3 and 1.45 M$_\odot$ models represented in \figurename{}~\ref{HR}. This mass range is typical of subgiant stars observed with {\it Kepler}. For all the considered models, the wave-related timescale was computed assuming a smooth rotation profile, \eqref{smooth domega}, and using the default values for the plume parameters as described in \sectionname{}~\ref{model parameters}. In all cases, we find a similar qualitative behavior to the one depicted in \figurename{}~\ref{timescale sg rg}. Whatever the mass, IGW can efficiently spin down all the region below the hydrogen-burning shell if the amplitude of the differential rotation is higher than a given threshold. This theoretical threshold \smash{$\overline{\Delta \Omega}_{{\rm th}}$} is displayed in \figurename{}~\ref{loop} as a function of $\log~g$ and the mass. 
   \begin{table}[]
      \caption[]{Characteristics of the six subgiant and early red giant stars studied by \cite{Deheuvels2014}. The mass were obtained with a fitting procedure using the evolution code CESTAM, $\log~g$ was determined by seismic scaling relations and the amplitude of the radial-differential rotation between the center of the star and the base of the convective zone was inferred from the OLA method.}
         \label{table2}
     $$ 
         \begin{array}{ccc}
           \hline
	  \hline
	\noalign{\smallskip}
            M/M_\odot& \log~g& \Delta \Omega_{{\rm obs}}~(\upmu\mathrm{rad~s}^{-1}) \\
	\noalign{\smallskip}
            \hline
	\noalign{\smallskip}
	1.22&3.83 \pm 0.04& 2.00 \pm 0.66\\
	\noalign{\smallskip}
	1.27&3.77\pm 0.02&2.92\pm0.79\\
	\noalign{\smallskip}
	1.14&3.76\pm0.04&4.65\pm 0.30\\
	\noalign{\smallskip}
	1.26&3.71\pm 0.03&9.04\pm0.46\\
	\noalign{\smallskip}
	1.39&3.68\pm 0.02&7.48 \pm0.26\\
	\noalign{\smallskip}
	1.07&3.60\pm 0.02&9.31\pm 0.31 \\
	\noalign{\smallskip}
	\hline
         \end{array}
     $$ 
   \end{table}
%
\subsection{Comparison with the observations}
The theoretical thresholds  are compared to the amplitude of the differential rotation observed in six {\it Kepler} subgiant and early red giant stars studied by \citet[see \tablename{}~\ref{table2}]{Deheuvels2014} and the upper value of the core rotation observed in more evolved red giant stars by \cite{Mosser2012a}. The influence of the mass on the threshold value, as well as the comparison with the observations, differ according to the evolutionary status.

\begin{figure*}[]
\centering
\includegraphics[scale=0.7,trim= 0.5cm 0cm 0cm 0cm, clip]{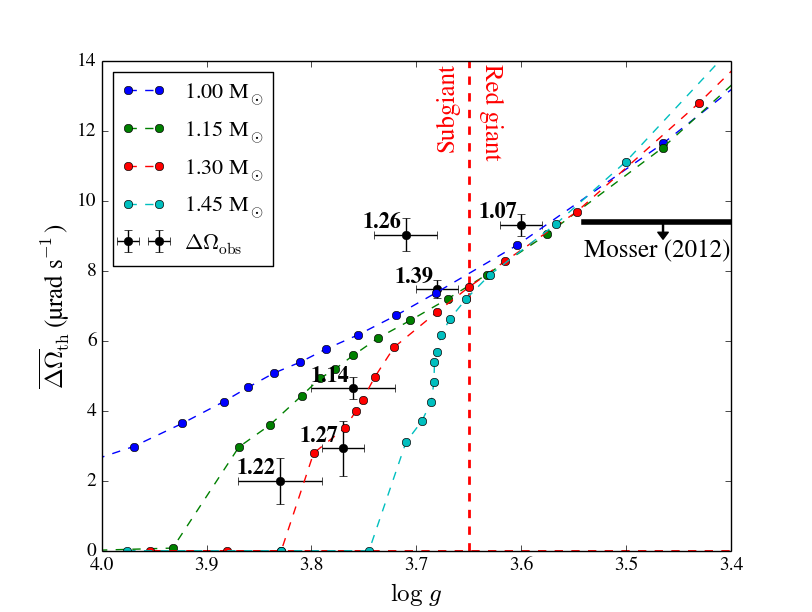}
\caption{Theoretical threshold for the amplitude of the radial-differential rotation as a function of the $\log$ $g$ for the considered evolutionary sequences. It is computed using the default values of the parameters given in \sectionname{}~\ref{model parameters} and assuming a smooth rotation profile as defined in \eqref{smooth domega}. The observations in subgiant stars, $\Delta \Omega_{{\rm obs}}$, are represented with error bars and masses in units of solar mass \citep[][see \tablename{}~\ref{table2}]{Deheuvels2014}. The vertical red dashed line symbolizes the transition between the subgiant phase to the beginning of the RGB. The maximum amplitude of the differential amplitude on the RGB as expected from the observations by \cite{Mosser2012b} is indicated by the horizontal thick black line.}
\label{loop}       
\end{figure*}

\paragraph{Red giant stars ($\log~g \gtrsim 3.55$).}
Figure~\ref{loop} shows that \smash{$\overline{\Delta \Omega}_{{\rm th}}$} is larger than about $10~\upmu$rad s$^{-1}$ in this mass range, which is well above the core rotation rates observed in red giants. We conclude that IGW are unable to affect, on their own, the internal rotation of these stars whatever the mass. Again, a large magnitude of the radiative damping is the cause. Moreover, \figurename{}~\ref{loop} shows that, at a given $\log$ $g$,  \smash{$\overline{\Delta \Omega}_{{\rm th}}$} converges toward the same value for all masses. Indeed, stars on the RGB are ascending the Hayashi line in the Hertzsprung-Russel diagram (see \figurename{}~\ref{HR}) and therefore, share similar properties at a given value of $\log$ $g$. Their internal structures are comparable, and so are the magnitude of the radiative damping, the contraction timescale and the convective plumes parameters $V_p$, $b/r_t$ and $\nu_p\sim \omega_c$. This is illustrated in \figurename{}~\ref{caract}. Therefore, angular momentum transport by IGW must have a similar influence on the internal rotation of these stars, so that \smash{$\overline{\Delta \Omega}_{{\rm th}}$} takes very close values whatever the mass for a given $\log$ $g$.

\paragraph{Subgiant and early red giant stars ($\log~g \lesssim 3.55$).}
Figure~\ref{loop} shows that the theoretical threshold for the differential rotation amplitude, \smash{$\overline{\Delta \Omega}_{{\rm th}}$}, is in good agreement with the observed amplitude of the differential amplitude in subgiant and early red giant stars. We find that \smash{$\overline{\Delta \Omega}_{{\rm th}} \lesssim 10~\upmu$rad s$^{-1}$} for these stars. Therefore, the conclusions made in \sectionname{}~\ref{t_w sg} hold for the considered mass range. Figure~\ref{loop} also shows that the higher the mass, the lower the theoretical threshold in these stars. Indeed, as shown in \figurename{}~\ref{caract}, the higher the mass, the higher the turnover convective frequency (and also $\nu_p$)  and the ratio $r_t/b$, so that, as demonstrated in \sectionname{}~\ref{uncertainties}, the more efficient the angular momentum transport by IGW. Therefore, the mass dependence of \smash{$\overline{\Delta \Omega}_{{\rm th}}$} at given $\log$ $g$ on the subgiant branch can be explained simultaneously by the increase in $\nu_p$ and $r_t/b$ as the mass increases.

\subsection{IGW-driven regulation of the core rotation in subgiant stars}

More than comparable orders of magnitude, the theoretical threshold values closely match the observations. Although the $1.26$~M$_\odot$ star slightly deviates from the theoretical curve, it is however not far enough to hinder the global agreement. Indeed, such a mismatch can be simultaneously explained by a small error in the estimate of the $\log~g$, a moderate decrease in $\nu_p$ or in $r_t/b$ (see \sectionname{}~\ref{uncertainties}), as well as uncertainties on stellar modeling. Thus, this agreement supports the idea that angular momentum transport by plume-induced IGW could operate as a regulating actor of the core rotation evolution in subgiant and early red giant stars. 

The scenario can be presented as follows. During the evolution on the subgiant branch, the internal structure of stars (or the magnitude of the radiative damping) and the characteristics of the convective plumes (i.e., $\mathcal{F}_0$, $\nu_p$ and $b$) imposes a threshold value for the differential rotation, \smash{$\overline{\Delta \Omega}_{{\rm th}}$}, above which plume-induced IGW can counteract the spin up of the innermost layers due to their contraction. As the core contracts, the wave radiative damping increases throughout the radiative zone, resulting in an increase in the value of the threshold over time (see \sectionname{}~\ref{t_w sg} and \appendixname{}~\ref{influence of damping}). Since the amplitude of the differential rotation obtained considering only local conservation of angular momentum is much higher than the observations \citep[e.g.,][see Fig. 8]{Ceillier2013}, we deduce that the increase in \smash{$\overline{\Delta \Omega}_{{\rm th}}$} occurs on a characteristic timescale much longer than the contraction timescale. Furthermore, in the limit of slow rotators, we can assume that rotation does not modify the equilibrium structure, so that variations in the instantaneous amplitude of the differential rotation \smash{$\overline{\Delta \Omega}$} have no consequences on \smash{$\overline{\Delta \Omega}_{{\rm th}}$} at a given evolutionary stage. In this framework, let us consider a star initially at the end of the main sequence with an amplitude of the differential rotation \smash{$\overline{\Delta \Omega}_0$}. At this stage, the system can evolve following two ways. If \smash{$\overline{\Delta \Omega}_0\le\overline{\Delta \Omega}_{{\rm th}}$}, then the core contraction dominates and \smash{$\overline{\Delta \Omega}$} increases with a timescale of $t_{{\rm cont}}$. Otherwise, IGW spin down the core of the star and \smash{$\overline{\Delta \Omega}$} decreases on characteristic timescales lower than the contraction timescale. Therefore, since the threshold value evolves on much larger timescales in both cases, \smash{$\overline{\Delta \Omega}$} becomes closer and closer to \smash{$\overline{\Delta \Omega}_{{\rm th}}$} in a first transient phase. After a reasonable time span, a stable steady state is reached in which the amplitude of the differential rotation is close to the threshold (i.e., \smash{$\overline{\Delta \Omega}\sim\overline{\Delta \Omega}_{{\rm th}}$}), independently of the initial conditions \smash{$\overline{\Delta \Omega}_0$}. The stability of the steady state is the consequence of the combined effects of the core contraction and of the angular momentum transport by IGW, which will counter any departure from this equilibrium. Finally, since the contraction timescale is much smaller than evolution timescale on the subgiant branch, the steady state is reached well before the beginning of the red giant branch. Therefore, the dynamical equilibrium between the core contraction and the transport by IGW holds over time so that \smash{$\overline{\Delta \Omega}$} follows \smash{$\overline{\Delta \Omega}_{{\rm th}}$} during the subgiant evolution, whence the observational agreement.

%
\section{Discussion}
%
\label{Discussion}

\begin{figure}[t]
\centering
\includegraphics[scale=0.5,trim= 0.5cm 0cm 0cm 0cm, clip]{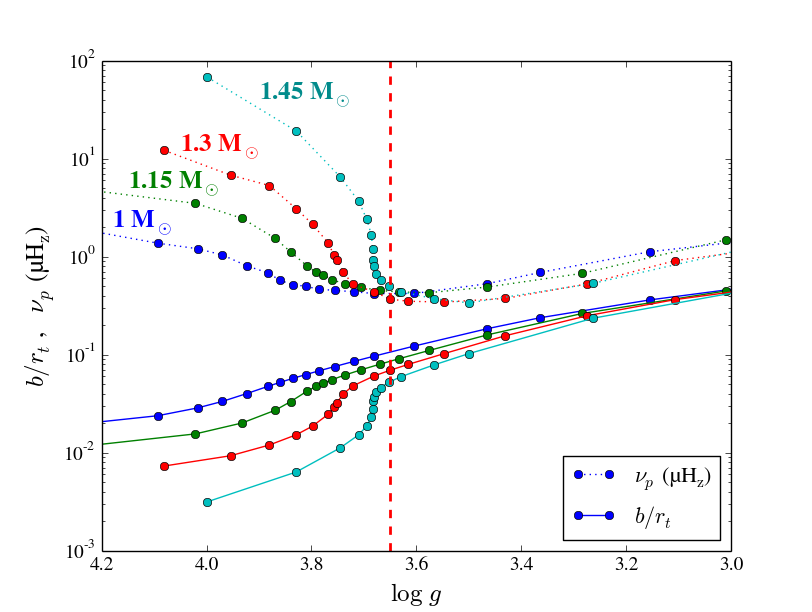}
\caption{Occurrence frequency (dotted lines) and width of the convective plumes normalized by the radius at the top of the radiative zone (solid lines), estimated at the base of the convective zone and plotted as a function of $\log$~$g$. The plume occurrence frequency, $\nu_p$, is assumed to be equal to the convective turnover frequency, $\omega_c$, which was estimated via the mixing-length theory. Each color is associated with a mass.}
\label{caract}       
\end{figure}

\subsection{Comparison with the transport by turbulence-induced IGW}

\cite{Fuller2014} followed a similar approach as we used in this work, but considering angular momentum transport by turbulence-induced IGW. They assumed that IGW are generated by convective motions in the overshooting region and used the excitation model proposed by \cite{Lecoanet2013}. As a result, they showed that the evolution of the helium core rotation should decouple from the influence of convectively-excited IGW when stars reach about the middle of the subgiant branch. As explained in their paper and discussed in \appendixname{}~\ref{wave radiative damping}, this is the consequence of the increase in the wave radiative damping as the innermost layers contract. However, our conclusions differ from their results since we observe that the decoupling happens later, at the beginning of the ascent of the RGB. This difference can be explained by the two main following points.

Firstly, the amplitude and the shape of the wave excitation spectrum considered in both studies are different. The turbulence-induced wave spectrum is a decreasing power law with respect to frequency whereas the plume-induced one has a Gaussian profile. Moreover, turbulence-induced IGW have most of their energy in low angular degrees following the model of \cite{Lecoanet2013}. Since the wave radiative damping is in addition proportional to~$l^3$, \cite{Fuller2014} considered $l\sim 1$, $|m|=1$ wave components only. Conversely, plume-induced IGW are preferentially excited at higher angular degrees. Indeed, the size of the plumes is three to ten times smaller than the size of the biggest convective eddies as described by the MLT along the evolutionary tracks that we have considered. Therefore, the effect of angular degrees $l\gtrsim 2$ must be taken into account in this case.

Secondly, Fuller et al's diagnosis is based on the comparison of the contraction timescale with the characteristic timescale on which IGW can modify the total angular momentum inside a sphere of radius $r$. It is therefore representative of the mean effect induced by the transport by IGW on each mass shell over this region and does not account for local effects. The authors argued in favor of this timescale because it does not depend on the rotation profile below the radius $r$, which is generally unknown. However, we stress that it depends on the rotation profile above the radius $r$. Indeed, the differential rotation creates the asymmetry between prograde and retrograde waves and operates like a wave filter via the wave damping in \eqref{damping} (see the difference between the results obtained considering either a smooth or a sharp rotation profile in \figurename{}~\ref{timescale var}). Furthermore, we have seen in \sectionname{}~\ref{IGW transport} and \sectionname{}~\ref{regulation} that its influence on the transport by IGW is predominant in the innermost layers of stars. Therefore, we preferred to use the local wave-related timescale, as defined in \eqref{t_w}, and assumed a given rotation profile. To compensate our lack of knowledge about the rotation profile, we considered a smooth and a sharp shape for the profile. Although both approaches are complementary, the one used in this work has the advantage of taking into account the amplitude of the differential rotation whose value can be constrained by seismic observations.

In this framework, it is worth investigating the efficiency of the angular momentum transport by turbulent-induced IGW using the comparison between the contraction timescale and the local wave-related timescale as given in \eqref{t_w}. We considered four models on the subgiant branch with $\log~g \sim 3.75$ and masses of 1,1.15, 1.3 and 1.45 M$_\odot$. We used the generation model by turbulent pressure as described in \cite{Kumar1999}, with wavefunctions estimated in the WKB approximations, and considered a smooth rotation profile as given in \eqref{smooth domega}. Finally, we found that turbulence-induced IGW can counteract the spin up of the region below the hydrogen-burning shell if the amplitude of the differential rotation is higher than 12, 10, 8 and 6~$\upmu$rad~s$^{-1}$ for 1, 1.15, 1.3 and 1.45 M$_\odot$, respectively. These values are well above observations at this value of $\log$~$g$. This result is in agreement with the conclusions of \cite{Fuller2014}. They are also higher than the threshold amplitudes obtained by considering IGW generated by penetrative convection. Therefore, we can conclude that plume-induced IGW are more efficient than turbulence-induced ones on the subgiant branch. Unlike plume-induced IGW, turbulence-induced IGW seem unable, on their own, to explain the observations in subgiant stars.

\subsection{Extraction mechanism and influence of the differential rotation amplitude in the inner radiative region}

In the theoretical framework used in this work, IGW and rotation are intricately related. The role of the amplitude of the differential rotation on the transport by IGW has been shown to be essential. Nevertheless, the interpretation of the behavior of $t_{{\rm w}}$ with \smash{$\overline{\Delta \Omega}$} in \sectionname{}~\ref{t_w sg} is not straightforward. Indeed, following \eqref{total flux}, the angular momentum transport by IGW is the consequence of the competition between three main physical ingredients: the magnitude of the radiative diffusion, the excitation wave spectrum and the oscillation frequency Doppler-shift. In order to grasp the extraction mechanism by IGW, a semi-analytical toy model has been developed and is described in \appendixname{}~\ref{model appendix}. Hereafter, we summarize the main points learned from the toy model. 

The efficiency of the angular momentum transport induced by IGW in \eqref{domega_dt} is represented by the divergence of the wave flux, $\dot{J}_{{\rm w}}$. This latter can be approximated by \smash{$\dot{J}_{{\rm w}}(r) \approx \tau(r) \mathcal{F}_{r,w}(r_t) e^{-\tau (r)}/r$}, with $\mathcal{F}_{r,w}$ given by \eqref{total flux}, $\tau(r)~\sim~r/L_D(r)$ and $L_D$ the local radial wave damping lengthscale (see \appendixname{}~\ref{wave radiative damping} for details). Therefore, the radiative damping must be neither too high nor too low for $\dot{J}_{{\rm w}}$ to be high enough. Consequently, the balance is reached and IGW are efficiently absorbed into the medium if the wave radiative damping in \eqref{damping}, or the acoustic damping depth, is close to unity (i.e., $\tau \sim 1$). From this statement, we can deduce several features of the transport by IGW whittled down to the four following points:
\begin{enumerate}
\item Since the wave frequency Doppler-shift is quasi null at the top of the radiative zone (i.e., $\delta \Omega(r_t) \approx 0$), very low-frequency high-degree IGW rapidly satisfy $\tau \sim 1$ as they travel downwards. They are then absorbed just below the base of the convective zone. As mentioned in \sectionname{}~\ref{t_w rg}, the wave-related timescale is very low in this region. IGW are thus able to modify the rotation rate on very short timescales.

\item Deeper in the star, the retrograde IGW ($m<0$) satisfying $\tau\sim 1$ have lower frequencies $\omega$ than progade ones ($m~>~0$) at given $|m|$ and $l$. Since the excitation wave spectrum is a decreasing function of $\omega$, retrograde IGW deposit more angular momentum than prograde ones. As a result, the net wave flux of angular momentum is negative, $\dot{J}_{{\rm w}}>0$ in \eqref{domega_dt} and IGW can therefore counteract the spin up due to the core contraction.

\item At given $l$, $|m|$ and $\omega$, an increase in $\Delta \Omega$ leads to a decrease in the wave radiative damping in \eqref{damping} for the retrograde IGW that can therefore go deeper into the star. In other words, in a given layer, the higher $\Delta \Omega$, the lower the wave frequencies absorbed into the medium and thus the higher the momentum transferred into the mean flow. As a consequence, an increase in the amplitude of the differential rotation results in an increase in the efficiency of the transport by IGW, and so a decrease in the wave-related timescale in deep layers of the radiative zone.
 
\item The absorption condition $\tau \sim 1$ in the helium core can be satisfied by IGW with high angular degrees on the condition that the Doppler-shift $|m|\Delta \Omega$ is large enough and that their frequencies $\omega$ are high enough to overcome the damping at the top of the radiative zone (where $\delta \Omega(r) \approx 0$). Moreover, since the wave excitation spectrum \eqref{wave energy flux} is maximum for $l_{{\rm max}}~\sim~r_t/b \gg1$, high-$l$
high-$|m|$ retrograde IGW can be responsible for the larger deposit of angular momentum in the helium core. This is what is observed in the subgiant models considered in this work when the amplitude of the differential rotation is of the order of magnitude of the threshold value (see also discussions in \sectionname{}~\ref{hypotheses transport} and \appendixname{}~\ref{absorbed component}). This statement goes against the commonly adopted view in which only low-degrees IGW can affect the rotation of the inner layers because of the wave damping depending on $l^3$.
\end{enumerate}
Therefore, simple arguments based on the condition of absorption given by $\tau \sim 1$ enable us to understand the influence of the differential rotation amplitude on the efficiency of the transport by IGW. It is also worth mentioning the high sensitivity of the wave-related timescale in the helium core to the value of the differential rotation amplitude. An explanation is given in \appendixname{}~\ref{sensitivity}. Such features are consistent with the behavior observed in \sectionname{}~\ref{IGW transport}. We refer to \appendixname{}~\ref{model appendix} for a more thorough and technical analysis.

\subsection{Uncertainties related to the generation of IGW in evolved stars}

\label{limitations excitation}

The modeling of the transfer of energy from convective plumes into waves relies on simplifying assumptions. Here, we briefly address the main issues concerning the generation model and refer to \cite{Pincon2016} for a more detailed discussion. 

As a first check, we found that the estimate of the wave energy flux induced by penetrative convection at the base of the convective zone is lower than 1\% of the stellar energy flux for all the considered models, that is well lower than the plume kinetic energy flux. The assumption of no feedback from the waves on the plumes in the penetration region is thus verified, as well as the conservation of energy. 

Following \eqref{wave energy flux}, the model also depends on several parameters. Considering a wide range of values for the model parameters, we have shown in \sectionname{}~\ref{uncertainties} that the uncertainties related to their estimate do not qualitatively modify the conclusions made from the timescales comparison in \sectionname{}~\ref{comparison}.

Lastly, the generation model neglects the effect of the Coriolis force on the waves in the excitation region. This hypothesis is valid for wave frequencies verifying $\omega \gtrsim 2\Omega(r_t)$, with $\Omega(r_t)$ the rotation rate at the top of the radiative zone.
Near the top of the radiative zone, \citet[]{Deheuvels2014} found rotation rates going from about 1 to $0.5~\upmu$rad~s$^{-1}$ for values of $\log~g$ going from 3.85 to 3.6 on the subgiant branch. This is in the same order of magnitude as the convective turnover frequency, and so, by assumption, as the plume occurrence frequency $\nu_p$ (see \figurename{}~\ref{caract}). Therefore, a non-negligible part of the wave excitation spectrum at low frequencies, say for $\omega$ below more or less $1~\upmu$rad~s$^{-1}$ for all the considered models, is affected by rotation. However, the low-frequency waves such as $\omega \lesssim 2\Omega(r_t)$ are expected to be absorbed just below the base of the convective zone, so that only the uppermost layers of the radiative zone are concerned by the uncertainties related to this point. Indeed, as discussed in \sectionname{}~\ref{hypotheses transport}, this range of low frequencies is not responsible for the deposit of angular momentum below the hydrogen-burning shell and, therefore, has no effect on the results obtained in this region.

\subsection{Influence of the Coriolis force on the propagation of IGW}
\label{hypotheses transport}

For the sake of simplicity, the modeling of the angular momentum transport by IGW and the derivation of \eqref{total flux} also neglect the effect of the Coriolis force on the propagation of the waves \citep[e.g.,][]{Mathis2009}, as well as the wave heat flux \citep[e.g.,][]{Belkacem2015b,Belkacem2015a}. Such approximations remain valid if the wave intrinsic frequency with respect to the co-rotating frame is higher than twice the rotation rate, or in other words, $\omega-m\delta \Omega(r) \gtrsim 2\Omega(r)$. While this is justified in the case of mixed modes with high oscillation frequencies, this is verified neither for waves near their critical layers (i.e., where $\omega=m\delta \Omega$), nor for low-frequency waves. However, it turns out a posteriori that the retrograde IGW ($m<0$) that efficiently deposit angular momentum into the medium below the hydrogen-burning shell, when the differential rotation amplitude is equal to about the threshold \smash{$\overline{\Delta \Omega}_{{\rm th}}$}, respect the condition $\omega+|m|\delta \Omega(r) \gtrsim 2\Omega(r)$ throughout the radiative zone, or equivalently $\omega+(|m|-2)\delta \Omega(r) \gtrsim 2\Omega(r_t)$ whatever $r$. Indeed, at the very beginning of the subgiant branch where \smash{$\overline{\Delta \Omega}_{{\rm th}}\lesssim \nu_p$}, we find that retrograde IGW such as $\omega \gtrsim 7~\upmu$rad s$^{-1}$ and $l\sim|m|\sim 1$ are mainly responsible for the deposit of angular momentum below the hydrogen-burning shell. Later on the subgiant branch where \smash{$\overline{\Delta \Omega}_{{\rm th}} \gtrsim \nu_p$}, we find this time that retrograde IGW such as $2\lesssim\omega \lesssim 6~\upmu$rad s$^{-1}$ and $l\sim |m|\gtrsim 20$ efficiently transport angular momentum into the helium core. Such high azimuthal numbers are possible because the wave energy flux generated at the base of the convective zone, \eqref{wave energy flux}, is maximum for $l_{{\rm max}} \sim r_t/b \gg 1$ (see \figurename{}~\ref{caract}). Such types of behavior are confirmed and discussed through the toy model in \appendixname{}~\ref{key points}. Moreover, since the rotation rate at the top of the radiative zone, $\Omega(r_t)$, is about or lower than $1~\upmu$rad s$^{-1}$ over the subgiant branch from the observations by \cite{Deheuvels2014} (see discussion in \sectionname{}~\ref{limitations excitation}), we checked in all the considered cases that these wave components satisfy $\omega+(|m|-2)\delta \Omega(r) \gtrsim 2\Omega(r_t)$ whatever $r$. This suggests that the results of the timescales comparison in the helium core of the considered models is only slightly impacted by the effect of the Coriolis force. Therefore, although further work is needed to properly tackle the issue, neglecting the Coriolis force seems reasonable for a first estimate.

%
\section{Conclusions and perspectives}
%
\label{Conclusion}

In this work, we explored the efficiency of the angular momentum transport induced by IGW from the subgiant branch to the beginning of the red giant branch of low-mass stars. We considered several models chosen at different evolutionary stages and assumed a given internal rotation profile for each of them. The wave flux of angular momentum was estimated using the semi-analytical generation model by penetrative convection as proposed in \cite{Pincon2016}. As a result, the local timescale associated with the transport induced by IGW, $t_{{\rm w}}$, was computed and compared to the contraction timescale, $t_{{\rm cont}}$, throughout the radiative zone of the considered models. 

We found that IGW can counteract the spin up due to the contraction of the layers throughout the helium core (i.e., $t_{{\rm w}} < t_{{\rm cont}}$) when the amplitude of the radial-differential rotation between the center of the star and the top of the radiative zone is higher than a threshold value. Indeed, an increase in the differential rotation amplitude lowers the radiative damping of retrograde IGW that can thus reach deeper layers and extract more angular momentum from the helium core. In subgiant stars, we obtained theoretical thresholds for the radial-differential rotation amplitude that are consistent with the values of the differential rotation observed by \cite{Deheuvels2014}. We also found that the angular momentum transport by plume-induced IGW is more efficient than the one by turbulence-induced IGW. We then conclude that plume-induced IGW are able to hinder, on their own, the establishment of a strong differential rotation driven by the core contraction along the evolution on the subgiant branch.
Unlike subgiant stars, the theoretical thresholds obtained in red giant stars turned out to be well above the core rotation rates inferred by \cite{Mosser2012a}. This mainly results from an excessive magnitude of the radiative damping in these stars \citep[see also][for turbulence-induced IGW]{Fuller2014}. This discrepancy leads to the conclusion that IGW are unable, on their own, to affect the innermost rotation of red giant stars. An additional mechanism must be found to efficiently spin down their core. In both cases, the results remain valid under the assumption of a smooth rotation profile between the base of the convective zone and the center of the star. We also checked that these conclusions qualitatively hold for a wide range of values considered for the excitation model parameters. 

The theoretical thresholds for the radial-differential rotation amplitude were shown to match in a noteworthy way the observations in the six {{\it Kepler} subgiant and early red giant stars studied by \cite{Deheuvels2014}. Such similarities are interpreted as the result of a regulation process driven by plume-induced IGW during the subgiant stage. In this scenario, the combined effects of the core contraction (inducing a spin up of the core) and of the transport by IGW (inducing a spin down of the core) lead to a stable steady state in which the radial-differential rotation amplitude remains close to the threshold values (see \sectionname{}~\ref{regulation} and \figurename{}~\ref{loop}). If such a wave-driven process is further confirmed, it will have several consequences. First, the existence of a sharp rotation profile, with a strong gradient in the vicinity of the hydrogen-burning shell, will be excluded since it prevents IGW from reaching the innermost layers of stars.
Second, the increase in the number of observations in subgiant stars will provide a diagnosis to constrain the parameters of the generation model by penetrative convection. Indeed, the matching between the observed amplitudes of the differential rotation and the theoretical thresholds will bring information on the plume parameters and will give us the interesting opportunity to indirectly probe the convective plumes at the base of the convective zone along the subgiant branch. Even though this interpretation is based on a comparison with only six observed subgiant stars, such a regulation mechanism is promising and stresses out the major role that IGW can play in the rotation history of stars. Additional observational constraints, as well as more exhaustive computations are needed in future to asses the relevance of this process.

The study performed in this work demonstrates that the angular momentum transport by plume-induced IGW in subgiant and early red giant stars is efficient and calls for more sophisticated investigations. The next step is to properly implement the transport by IGW generated by penetrative convection with the other transport mechanisms in a 1D stellar evolution code. While the timescales comparison as it is used in this work gives a hint about the global influence of IGW on the innermost layers of stars, the resulting conclusions ignore effects induced by IGW in very localized regions of the radiative zone, as well as the coupling between the various transport processes. For instance, we do not exclude that IGW that are responsible for the large deposit of angular momentum above the hydrogen-burning shell in red giant stars ($t_{{\rm w}} / t_{{\rm cont}}\ll1$ in \figurename{}~\ref{timescale sg rg}) could interact with another transport process, for instance the meridional circulation, and could indirectly boost the extraction of angular momentum from the core. As another example, we have shown that IGW can efficiently slow down the rotation mean flow in the vicinity of the center in the subgiant stars ($t_{{\rm w}} / t_{{\rm cont}}\ll1$ in \figurename{}~\ref{timescale sg rg}, see also \appendixname{}~\ref{behavior at the center} for explanations). This trend has already been observed in previous numerical computations \citep[e.g.,][]{Talon2005,Charbonnel2013,Mathis2013}. We can thus wonder how this could influence the global evolution over time of the rotation profile throughout the radiative zone of subgiant stars.
Modeling the dynamical evolution of the rotation profile with the interactions between all the transport processes along stellar lifetimes is needed to answer such questions and is a necessary step toward a full and complete understanding of the rotation history of stars.

On a long term, investigations will have to go beyond the simplifying assumptions used for the description of the convective plumes and the propagation of internal waves. The influence of the interaction with the upflow and of the sphericity in evolved stars will have to be included in the estimate of the plume parameters. Further theoretical efforts are also necessary to describe the plume destruction process, which remains still poorly known. In this framework, 3D numerical simulations of extended convective envelopes with more realistic Prandtl and Reynolds numbers (or in a regime from which we can scale to the realistic dimensionless numbers in stars) would help to constrain the plume parameters. Moreover, the coupling between IGW and shear-induced turbulence near the critical layers \citep[see][for an exhaustive description]{Alvan2013} and the effect of the over-transmission or reflection processes on the transport by IGW will have to be investigated. Also, the contribution of the wave heat flux to the angular momentum transport equation \citep[e.g.,][]{Belkacem2015a}, as well as the effect of the Coriolis force and of the rotation gradient on the propagation of the waves \citep[e.g.,][]{Mathis2009}, will have to be properly taken into account. Undertaking a global study of the transport by internal waves in stellar radiative interiors that tackles with the abovementioned issues is challenging since it requires a 2D description of the internal waves and of the plume dynamics throughout an extended convective envelope.


\begin{acknowledgements}
We acknowledge Yveline Lebreton for the fruitful discussions about modeling with the evolution code CESTAM and the computation of the models used in this work. CP is especially grateful for her friendship and her support since the undergraduate level.
\end{acknowledgements}

%
\bibliographystyle{aa}
\bibliography{bib}
%

\appendix

\section{Wave radiative damping}
\label{wave radiative damping}

In the following, we describe the behavior of the wave damping throughout the radiative zone of stars as well as its evolution over time. Guided by stellar models, we propose a simple modeling that enables us to investigate the angular momentum transport by IGW in a tractable way. This model will be used in \appendixname{}~\ref{model appendix}. 

\subsection{Spatial behavior and time evolution}
\label{spatial behavior}

To discuss the behavior of the wave damping, it is convenient to define the local radial damping lengthscale of the wave energy, $L_D$. It is defined by
\begin{align}
L_D (r,\hat{\omega},l) &\approx \frac{\hat{\omega}^4 r^3}{[l(l+1)]^{3/2} K N^2 N_T} \mbox{ ,}
\label{damping length}
\end{align}
so that \eqref{damping}, or the damping acoustic depth, can be rewritten 
\begin{align}
\tau(r, \hat{\omega},l)&=\int_r^{r_t} \frac{{\rm d} r^\prime}{L_D(r^\prime, \hat{\omega},l)}\mbox{ ,} \label{tau}
\end{align}
with $\hat{\omega}$ the wave intrinsic frequency defined in \eqref{intrinsic frequency}, $r$ the radius, $l$ the wave angular degree, $K$ the radiative diffusivity, $N$ ($N_T$) the Brunt-Väisälä frequency with (without) the gradient of the chemical composition, and $r_t$ the radius at the top of the radiative zone.
$L_D$ is plotted in \figurename{}~\ref{damping sg rg} for the subgiant and the red giant 1M$_\odot$ models that are considered as examples in \sectionname{}~\ref{comparison}. For both models, the radial damping lengthscale decreases by more than five orders of magnitude from the top of the radiative zone to the hydrogen-burning shell where it is minimum.
A value of the ratio $L_D/r$ much lower than unity indicates that $\tau\gg 1$. This means that the wave is locally absorbed into the medium and cannot reach deeper layers. In both examples displayed in \figurename{}~\ref{damping sg rg}, the considered wave component, characterized by $l=1$ and $\hat{\omega}=1~\upmu$rad~s$^{-1}$, is considerably dissipated immediately after the top of the radiative zone. Nevertheless, following \eqref{damping length}, an increase in the intrinsic wave frequency (i.e., following \eqref{intrinsic frequency}, an increase in $\omega$ or in $-m\delta \Omega$) leads to a increase in $L_D$ such that the wave can go deeper into the star. In turn, a value of $L_D/r$ much higher than unity means that the wave will not be damped enough to locally deposit momentum into the medium. Figure~\ref{damping sg rg} also shows that $L_D$ is more than two orders of magnitude lower for a red giant than for a subgiant star. Indeed, as the core starts contracting at the end of the main sequence, the Brunt-Väisälä frequency increases in the radiative zone and the radius of a given mass shell decreases, resulting in an important decrease in \eqref{damping length}. Incoming waves are therefore more rapidly damped in red giant than in subgiant stars for a given value of the wave intrinsic frequency. We refer the reader to the discussion in \appendixname{}~\ref{influence of damping} for more details about the impact of the radiative damping on the transport by IGW.

To derive \eqref{damping length}, we note that we have supposed that the factor $N/(N^2-\hat{\omega}^2)^{1/2}$ in the integrand of \eqref{damping} is always close to unity. This is true far away from a reflexion point (i.e., where $N^2 = \hat{\omega}^2$). When $\omega^2 \sim N^2$, we can show by a first-order expansion that the integral \eqref{damping} does not diverge and that the contribution near this point to the total integral remains small. Therefore, waves are not dissipated into the medium near a reflection point, unlike at critical layers (i.e., where $\hat{\omega}=0$).

\begin{figure}[]
\centering
\includegraphics[scale=0.5,trim= 0.5cm 0cm 0cm 0cm, clip]{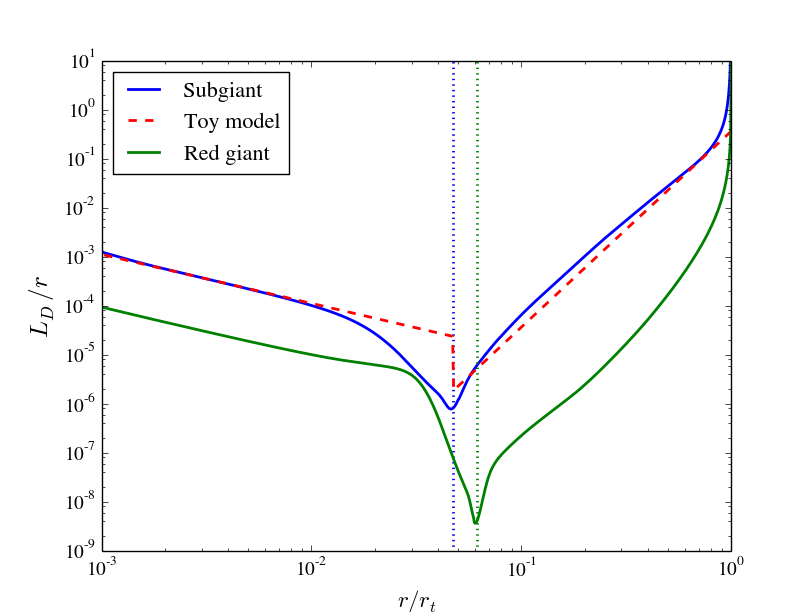}
\caption{Ratio of the characteristic radial damping lengthscale of the wave energy as given in \eqref{damping length} to the radius for both 1M$_\odot$ models considered in \sectionname{}~\ref{comparison}. It is plotted as a function of the radius normalized by its value at the top of the radiative zone, $r_t$, for a degree $l=1$ and an intrinsic wave frequency $\hat{\omega}=1~\upmu$rad~s$^{-1}$. The vertical dotted lines localizes the hydrogen-burning shell in both models. The profile assumed in \appendixname{}~\ref{basic modeling} is represented by the red dashed line for the subgiant model.}
\label{damping sg rg}      
\end{figure}

\subsection{An approximate for $L_D$}
\label{basic modeling}
Estimates of $L_D$ for stellar models show that it is possible to approximate \eqref{damping length} with the help of power laws. We can then assume that
\begin{align}
L_D&\approx \left \lbrace
\begin{array}{cl}
\dfrac{\hat{\omega}^4 r_t^3}{[l(l+1)]^{3/2} K_t N_t^3} \left( \dfrac{r}{r_t}\right)^{\beta}  &  \mbox{if   $r_{{\rm HS}} < r < r_t$,}\\
\dfrac{\mathcal{R}}{[l(l+1)]^{3/2}} \left( \dfrac{\hat{\omega}}{1~\upmu{\rm rad~s}^{-1}}\right)^4 & \mbox{if   $r<r_{{\rm HS}},$}
\end{array}
\right.
\label{damping length assumed} 
\end{align}
where the subscript $t$ refers to quantities evaluated at the top of the radiative zone, 
$\beta$ is a rational number,
$r_{{\rm HS}}$ is the radius at the hydrogen-burning shell 
and $\mathcal{R}$ is a constant so that $L_D(r_{{\rm HS}}^-)\ge~L_D(r_{{\rm HS}}^+)$. We note that \eqref{damping length assumed} does not take the gradient in the chemical composition into account, so that $N=N_T$.
All these quantities can be provided by a stellar model.
A comparison with the 1M$_\odot$ subgiant model, considered in \sectionname{}~\ref{comparison}, is plotted in \figurename{} \ref{damping sg rg}. For our purpose, the discontinuity in the modeling at the hydrogen-burning shell does not matter. Indeed, since $L_D$ sharply increases below the hydrogen burning-shell, the contribution of this region to the total wave damping integral is negligible at first approximation compared to the one of the region around the hydrogen-burning shell. The useful physical quantities extracted from the model are listed in \tablename{}~\ref{table1}. This set of parameters will be used as inputs for the toy model presented in \appendixname{}~\ref{model appendix}.
\renewcommand{\arraystretch}{1.4}
   \begin{table}
      \caption[]{Parameters extracted from the 1 M$_\odot$ subgiant model considered in \sectionname{}~\ref{comparison}, with $\log g=3.86$ and $\log L/L_\odot= 0.35$. }
         \label{table1}
     $$ 
         \begin{array}{cc|cc}
           \hline
	  \hline
           \multicolumn{4}{c}{1~{\rm M}_\odot~{\rm subgiant~model~parameters}} \\
            \hline
           N_t/2\pi  & 85~\upmu {\rm Hz}& \beta & 5   \\
	   r_t & 6.6~10^{8}~{\rm m}&   r_t / b & 19.2	\\
	  r_{{\rm HS}} & 3.1~10^7~{\rm m} & \overline{\Delta \Omega}_{{\rm th}}/2\pi & 780~{\rm nHz}\\
 	  K_t & 2.8~10^3~{\rm m}^2~{\rm s}^{-1} & \nu_p\sim\omega_c & 0.6 ~\upmu{\rm Hz} \\
	   \mathcal{R} & 2.1~10^{3}~{\rm m} & V_p & 174~{\rm m~s}^{-1}\\
	  A_{l=1} & 3.4~10^{-11}& N_t/ \overline{\Delta \Omega}_{{\rm th}}& 109\\
	  \hline
         \end{array}
     $$ 
   \end{table}
%

\subsection{Transport by IGW near the center ($r \lesssim r_{{\rm HS}}$)}
\label{behavior at the center}

Without going in more details, it is already possible to deduce from the description given in \sectionname{}~\ref{basic modeling} how the transport by IGW behaves in the vicinity of stellar centers. Indeed, since $r_{{\rm HS}} \ll r_t$ in evolved stars, we can assume that $\delta \Omega$ has already reached its maximum amplitude $\Delta \Omega$ when $r\lesssim~r_{{\rm HS}}$, so that $\hat{\omega}~\approx~\omega~-~m\Delta \Omega$. Therefore, following \eqref{damping length assumed}, $L_D$ can be supposed to be constant for $r \lesssim~r_ {{\rm HS}}$.
As a result, \eqref{tau} can be rewritten as
\begin{align}
\tau(r, \hat{\omega},l)\approx \frac{[l(l+1)]^{3/2}}{\hat{\omega}^4} \frac{(r_{{\rm HS}}-r)}{\mathcal{R}} + \int_{r_{{\rm HS}}}^{r_t} \frac{{\rm d} r^\prime}{L_D(r^\prime, \hat{\omega},l)}\mbox{ ,}
\label{tau center}
\end{align}
provided that $\hat{\omega} >0$. Using \eqref{total flux}, we thus find that \eqref{J_dot} is almost proportional to $1/r^2$ when $r \lesssim r_{{\rm HS}}$. Therefore, taking $\rho \approx \rho(r=0)$ and $\Omega \sim \Omega(r=0)$ in this region, the wave-related timescale, \eqref{t_w}, becomes proportional to $r^4$. As a consequence, the influence of the transport by IGW on the rotation increases as $r\lesssim r_{{\rm HS}}$ and $r$ tends to zero, and the wave-related timescale can become much lower than the contraction timescale in the vicinity of the center. Such a trend is well reproduced by numerical computations (see \figurename{}~\ref{timescale sg rg}). It results from the fact that the moment of inertia of a mass shell behaves as $r^2$ as $r$ decreases, and that the wave flux increases as $1/r^2$ as it focuses toward the center. This result holds true on the condition that $\delta \Omega$ is constant in the vicinity of the center. 

To go further, a more detailed analysis of the wave spectrum and the amplitude wave flux requires accounting for the wave damping from the base of the convective zone to the hydrogen-burning shell and modeling the rotation profile. This will be subject to the next section.

\section{Angular momentum transport by IGW in a semi-analytical toy model}
\label{model appendix}
In this section, we present simplified analytical expressions of the wave damping integral (or the damping acoustic depth), $\tau$, and of the divergence of the angular momentum wave flux, $\dot{J}_{{\rm w}}$, between the hydrogen-burning shell and the base of the convective envelope.
The following developments rely on two basic assumptions.
Firstly, the differential rotation is supposed to linearly vary throughout the radiative zone of the star, 
\begin{align}
\delta\Omega(r)=\Delta \Omega \left( 1 -\frac{r}{r_t}\right)~~\mathrm{for}~~r\le r_t \mbox{ ,}\label{domega linear}
\end{align}
where $\Delta \Omega>0$ is the maximum amplitude and $r_t$ is the radius at the top of the radiative zone.
Secondly, the radial damping lengthscale of the wave energy is approximated by power laws as given by \eqref{damping length assumed}.
Such a toy model enables us to easily identify the important parameters of the transport by IGW and to investigate their influence on its efficiency. In particular, it highlights the role of the amplitude of the differential rotation and facilitates the interpretation of the results delivered in the main text.

\subsection{Wave damping integral above the hydrogen-burning shell ($r_{{\rm HS}} \lesssim r \le r_t$)}
\begin{figure*}[t]
\centering
\includegraphics[scale=0.48,trim= 0.5cm 0cm 1.5cm 0cm, clip]{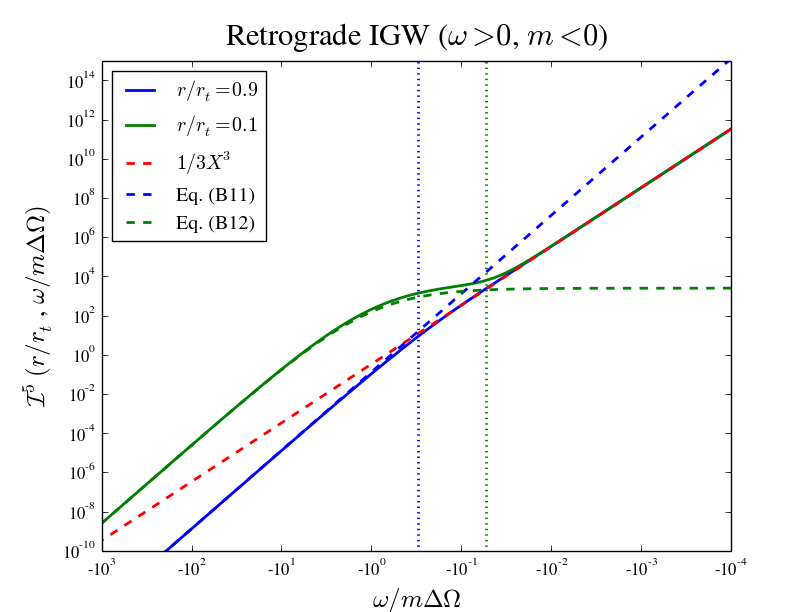}
\includegraphics[scale=0.48,trim= 0.5cm 0cm 1.5cm 0cm, clip]{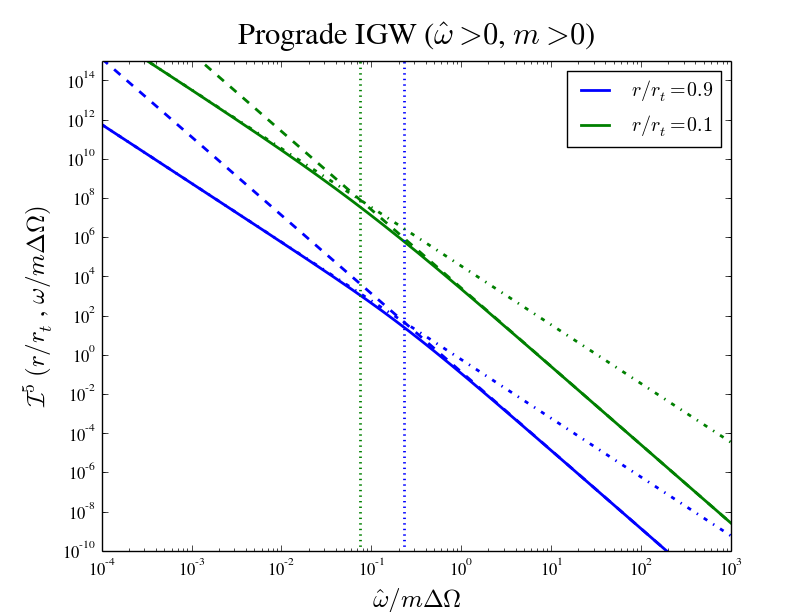}
\caption{{\bf Left:} Damping integral $\mathcal{I}^5(z,X)$ as defined in \eqref{integral I} for retrograde IGW ($m<0$) and as a function of $X=\omega/m\Delta \Omega$ (solid lines). The dashed lines represent the approximate expressions given in Eqs. (\ref{retrograde approximate BCZ}) and (\ref{retrograde approximate}). {\bf Right}: Damping integral $\mathcal{I}^5(z,X)$ for prograde IGW ($m>0$) as a function of $X-1+r/r_t=\hat{\omega}/m\Delta \Omega$ (solid lines). The dashed and the dashed-dotted lines respectively represent the approximate expressions at high and low values for $\hat{\omega}/m\Delta \Omega$ as given in \eqref{prograde approximate}. {\bf Both panels:} Two values of $z=r/r_t$ are considered, $r/r_t=0.9$ (blue color) and $r/r_t=0.1$ (green color). The vertical dotted lines indicates the values at the transition of regime, $X_t$, as defined in \appendixname{}~\ref{approximate I}.}
\label{comparaison I}       
\end{figure*}

The above assumptions make the computation of the wave damping integral possible in a semi-analytical way.
Using \eqref{domega linear} and \eqref{damping length assumed} for $r_{{\rm HS}}<r<r_t$, \eqref{tau} can be expressed as 
\begin{align}
\tau(r, \hat{\omega},l) &= A_l ~\chi_{|m|}^4 ~\mathcal{I}^\beta\left( \frac{r}{r_t}, \frac{\omega}{m\Delta \Omega} \right)\mbox{  ,}
\label{tau 2}
\end{align}
provided that $\omega > m \delta \Omega(r)$ and where we have defined
\begin{align}
A_l&=\frac{[l(l+1)]^{3/2} K_t}{r_t^2 N_t}\label{A_l}\\
 \chi_{|m|} &= \frac{N_t}{|m| \Delta \Omega}\label{chi m}\\
\mathcal{I}^\beta \left(z,X\right) &= \int_{z}^1 \frac{y^{-\beta}} {[X-(1-y)]^4} {\rm d} y \label{integral I} \mbox{ .}
\end{align}
If $\omega < m \delta \Omega(r)$, the waves have already been dissipated in upper layers (cf. critical layers). The damping process is thus controlled by two main parameters, $A_l$ and $\chi_{|m|}$. The first one is representative of the intensity of the radiative diffusion whereas the second one measures the frequency Doppler-shift.

\subsubsection{Exact solution for $\mathcal{I}^\beta(z,X)$}
\label{exact solution}

The solutions of \eqref{integral I} can be expressed in different ways depending on the exponent $\beta$ and the quantity $a = X-1$.
\paragraph{Case when $\beta \in \mathbb{N}$:} $\mathcal{I}^\beta$ is directly deduced by a recurrence relation. We found, following \cite{Gradshteyn2007},
\begin{align}
\mathcal{I}^\beta(z,X) =\frac{1}{1-\beta}\left(\left[ \frac{y^{1-\beta}}{a (a+y)^3}\right]_{z}^1+~\frac{\beta+2}{a}\mathcal{I}^{\beta-1}(z,X) \right) \mbox{  ,}
\end{align}
with the initial condition for $\beta=1$ given by
\begin{align}
\mathcal{I}^{1}(z,X) = \left[ \frac{11 a ^2 +15 a y + 6 y^2}{6 a^3(a +y)^3} -  \frac{ \ln|a+y|-\ln y}{a^4}\right]_z^1\mbox{  .}
\label{sol 1}
\end{align}

\paragraph{Case when $\beta~\notin \mathbb{N}$:} then, if $a \ge -1$, 
\begin{align}
\mathcal{I}^\beta(z,X) =
\left[ -\frac{_2F_1\left(\beta,\beta+3,\beta+4;\frac{a}{y+a}\right)}{(\beta+3)(y+a)^{\beta+3}}\right]_z^1
\mbox{  ,}
\label{sol 2}
\end{align}
which is also valid when $\beta$ is an integer, or otherwise
\begin{align}
\mathcal{I}^\beta(z,X) =
\left[ -y^{1-\beta}\frac{_2F_1\left(4,1-\beta,2-\beta;-\frac{y}{a}\right)}{a^4(\beta-1)}\right]_z^1\mbox{ ,}
\label{sol 3}
\end{align}
where $_2F_1$ is the hypergeometric function.
\\

As an illustration, the integral is displayed in \figurename{}~\ref{comparaison I} for $\beta=~5$ with $z=r/r_t$ and $X=\omega/m\Delta \Omega$. Two values of $z$ are considered. Negative values of $X$ correspond to retrograde IGW ($m<0$) and positive values of $X$ correspond to prograde IGW ($m>0$), on the condition that they have not been dissipated near their critical layers yet (i.e., such as $X>1-r/r_t$ or, similarly, $\hat{\omega}/m\Delta \Omega >0$).

\subsubsection{Approximate solutions for $\mathcal{I}^\beta$}
\label{approximate I}

We now search for approximate expressions of the wave damping integral. Considering different assumptions on the values of $z$ and $X$ leads us to distinguish three cases:
\begin{enumerate}

\item {\bf Retrograde IGW ($m<0,~X<0$) for $z\sim 1$:} 
\begin{align}
\mathcal{I}^\beta(z,X)\approx \left \lbrace
\begin{array}{cl}
\dfrac{1}{\beta -1}  \dfrac{z^{-\beta+1}-1}{|X|^4} &  \mbox{if   $|X| \gtrsim X_t = 3(1-z)$}\\
\dfrac{1}{3|X|^3} & \mbox{otherwise.} 
\end{array}
\right.
\label{retrograde approximate BCZ}
\end{align}
\item {\bf Retrograde IGW ($m<0,~X<0$) for $z \ll 1$:} 
\begin{align}
\mathcal{I}^\beta(z,X)\approx \left \lbrace
\begin{array}{cl}
\dfrac{1}{\beta -1} \dfrac{z^{-\beta+1}-1}{(|X|+1)^4}  &  \mbox{if   $|X| \gtrsim  X_t=\left(\dfrac{(\beta -1)/3}{z^{-\beta+1}-1}\right)^{1/3}$}\\
\dfrac{1}{3|X|^3} & \mbox{otherwise.} 
\end{array}
\right.
\label{retrograde approximate}
\end{align}
\item {\bf Prograde IGW ($m>0,~X > 1-z$) for $0<z<1$:} 
\begin{align}
\mathcal{I}^\beta(z,X)&\approx \left \lbrace
\begin{array}{cl}
\dfrac{1}{\beta -1}  \dfrac{z^{-\beta+1}-1}{(X-1+z)^4} &  \mbox{if   $X \gtrsim X_t= \dfrac{3(z-z^\beta)}{\beta -1} +1-z$}\\
\dfrac{1}{3} \dfrac{z^{-\beta}}{(X-1+z)^3}& \mbox{otherwise.}
\end{array}
\right.
\label{prograde approximate}
\end{align}

\end{enumerate}
In all cases, the values of transition, $X_t$, were obtained by equating both approximate expressions at small and large $|X|$ values. These approximations are validated in \figurename{}~\ref{comparaison I} by a comparison with the exact solution.

\subsubsection{Two regimes controlled by the Doppler-shift} 
\label{outcome damping}
In the three abovementioned cases, \figurename{}~\ref{comparaison I} as well as Eqs.~(\ref{retrograde approximate BCZ})-(\ref{prograde approximate}) show that the wave damping integral is composed of two regimes depending on whether the value of \smash{$|X|=\omega/|m|\Delta \Omega$} is greater than a transition value, denoted $X_t$, or not. This transition value depends on the radius in the star, and so on the distance already traveled by IGW. Above the transition $|X|>X_t$, the wave damping integral is dominated by the damping that IGW suffer in the vicinity of the layer at radius~$r$, which is far enough of their critical layers for prograde IGW. Below the transition $|X|<X_t$, the wave damping integral is dominated either by the damping they have suffered in the layers at the top of the radiative zone for retrograde IGW (i.e.,~where~$r\sim r_t$), or by the one they suffer near their critical layers for prograde IGW (i.e.,~where~$r\sim r_t[1-\omega/m\Delta \Omega]$).

Therefore, at a given radius in the star, the magnitude of the oscillation frequency Doppler-shift, $m \Delta \Omega$, selects the wave frequency at the transition between the two regimes. Figure
~\ref{comparaison I} and Eqs. (\ref{retrograde approximate BCZ})-(\ref{prograde approximate}) also show that a decrease in the radius~$r$, leading to an increase in the distance traveled by the waves, results in a decrease in the frequency at the transition. Indeed, the contribution of the local damping becomes dominant for a larger range of wave frequencies. 
We note that it also leads to a global increase in the damping integral for all frequencies, since the distance covered by the waves increases.

\subsection{Divergence of the wave flux $\dot{J}_{{\rm w}}$}
It remains to express the divergence of the angular momentum wave flux.
To do so, we assume that IGW are excited in a frequency range between 0 and $N_t$, with $N_t$ the value of the Brunt-Väisälä frequency at the top of the radiative zone.
Using Eqs.~(\ref{wave energy flux})-(\ref{damping}), the change of variable $\sigma=\omega/N_t$, and assuming a monotonously decreasing rotation profile, \eq{J_dot} can be rewritten in the following general form
\begin{align}
\dot{J}_{{\rm w}}(r)&=\sum_{l=1}^{+\infty} \sum_{m=1}^{+l} \left(\frac{\mathcal{F}_{l,|m|}(r)}{r} \int_{0}^{1} \mathcal{S}\left(\sigma;B\right)\mathcal{D}_{l,|m|}\left(\sigma,r\right)  \mathrm{d} \ln \sigma\right) 
\mbox{  ,}\label{J dot developed}
\end{align}
with
\begin{align}
\mathcal{F}_{l,|m|}(r)&=\frac{|m|}{\nu_p} \frac{r_t^2 }{r^2} 2\mathcal{F}_0 \sqrt{l(l+1)} e^{-\sqrt{l(l+1)} b^2/2 r_t^2}\\
\mathcal{S}\left(\sigma;B\right)&=e^{-\sigma^2/4 B^2}~~~~,~~~~B=\frac{\nu_p}{N_t}\label{spectrum func}\\
\mathcal{D}_{l,|m|}\left(\sigma,r\right)&=\widetilde{\mathcal{D}}_{l,+|m|}(\sigma,r)
-\widetilde{\mathcal{D}}_{l,-|m|}(\sigma,r)
\mbox{  ,}\label{damping term} 
\end{align}
where we have separated in \eqref{damping term} the retrograde ($-|m|$) and prograde ($+|m|$) parts and defined
\begin{align}
\widetilde{\mathcal{D}}_{l,m}\left(\sigma,r\right)&=H\left(\hat{\sigma}_m\right)\frac{r}{L_D }e^{-\tau} \label{D}\\
\hat{\sigma}_m(r)&=\frac{\hat{\omega}}{N_t}=\frac{\omega}{N_t} - m \frac{\delta \Omega (r)}{N_t} \mbox{  ,}\label{sigma chapeau}
\end{align}
and where $L_D$ is given by \eqref{damping length}. The Heaviside function, $H$, has been introduced to take into account that the prograde IGW that have already reached their critical layers (i.e such as $\hat{\omega} <0$) have been totally dissipated.

Using the assumptions made in the toy model, Eqs. (\ref{damping length assumed}), (\ref{A_l}) and (\ref{chi m}) for $r_{{\rm HS}} \le r \le r_t$, \eqref{D} can be formulated such as
\begin{align}
\widetilde{\mathcal{D}}_{l,m}(\sigma,r) &=A_l \chi_{|m|}^4 \left(\dfrac{r_t}{r}\right)^{\beta-1} \frac{e^{-\tau}}{\left[\sigma \chi_{|m|}-\mathrm{sgn}(m)~(1- r/r_t)\right]^4}\mbox{ ,}
\label{damping term 2}
\end{align}
with $\tau$ given by \eqref{tau 2}
\begin{align}
\tau&= A_l ~\chi_{|m|}^4 ~\mathcal{I}^\beta\left( ~\frac{r}{r_t}~,~\mathrm{sgn}(m)~ \sigma \chi_{|m|}~\right) \mbox{  ,}
\label{tau 3}
\end{align}
provided that $\sigma \chi_{|m|}>\mathrm{sgn}(m)~(1- r/r_t)$ and where ($\mathrm{sgn}$) is the sign function. At this stage, the set of equations Eqs. (\ref{tau center}), (\ref{tau 2})-(\ref{sol 3}) and (\ref{J dot developed})-(\ref{tau 3}) enables us to study in a tractable way the transport of angular momentum induced by IGW throughout the radiative zone of stars, as well as the role of the oscillation frequency Doppler-shift. 
As shown by \eqref{J dot developed}, the transport induced by IGW is the consequence of a balance between wave driving, through the $\mathcal{S}$ and $\mathcal{F}_{l,|m|}$ functions, and wave damping, through the damping term $\mathcal{D}_{l,|m|}$. While $\mathcal{S}$ and $\mathcal{F}_{l,|m|}$ are described in simple forms by the generation model, the behavior of the damping term $\mathcal{D}_{l,|m|}$ is much less easy to grasp. Moreover, $\dot{J}_{{\rm w}}$ depends on $\Delta \Omega$ through $\mathcal{D}_{l,|m|}$ only, so that a more detailed analysis of this term is necessary. 
For our purpose, we will focus in the following on the case $r/r_t \sim r_{{\rm HS}}/r_t \ll 1$. In other words, we will restrain the study to a region located around the hydrogen-burning shell, at the surroundings of the helium core.

\subsection{Damping term $\mathcal{D}_{l,|m|}$ near the H-burning shell} 

\subsubsection{Example with $l=|m|=1$}
\label{plot damping}
\begin{figure}
\centering
\includegraphics[scale=0.55,trim= 0.5cm 0cm 0cm 0cm, clip]{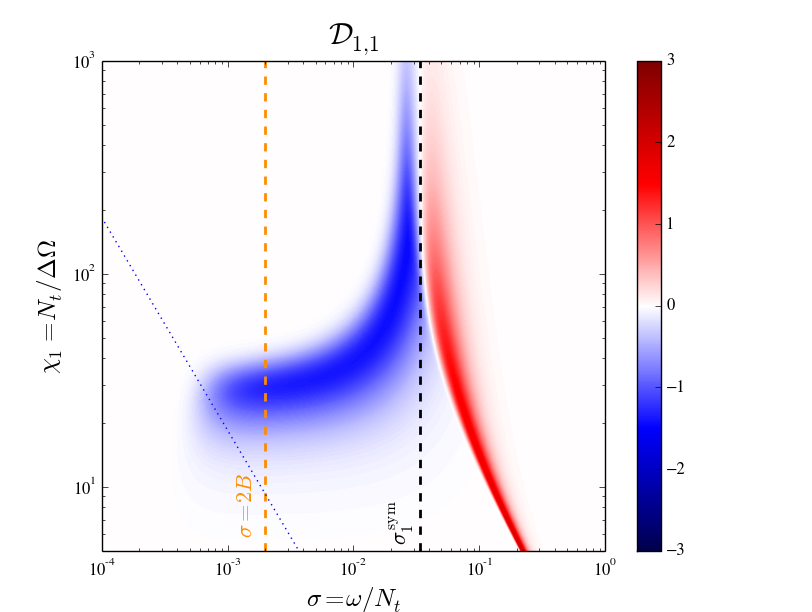}
\includegraphics[scale=0.5,trim= 0.5cm 0cm 0cm 0cm, clip]{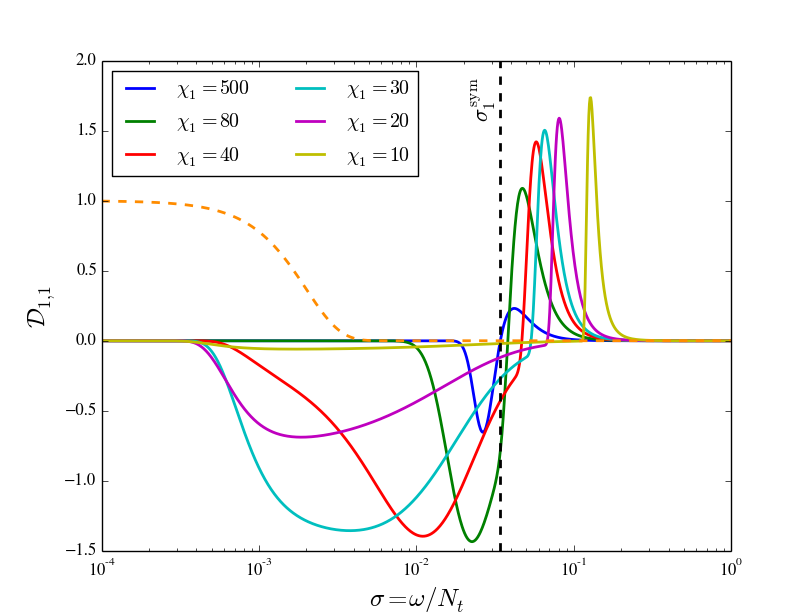}
\caption{{\bf Top:} Dimensionless damping term defined in \eqref{damping term} with $l=|m|=1$, as a function of $\sigma$ and $\chi_1$. It is plotted for the subgiant model parameters given in \tablename{}~\ref{table1} at a radius $r/r_t=0.05$, i.e., near the hydrogen-burning shell. The vertical orange dashed line corresponds to $\sigma=2 B$ as given in \eqref{spectrum func}. The vertical dark dashed line indicates the frequency at the separation between the retrograde and prograde parts of the damping term, $\sigma_1^{{\rm sym}}$, which was estimated using \eqref{sigma sym}. The dotted line represents the retrograde transition frequency as defined in \eqref{transition retro}. {\bf Bottom:} Cut sections for different values of $\chi_{1}$. The orange dashed line represents the Gaussian wave frequency spectrum defined in \eqref{spectrum func}. The vertical dark dashed line has the same signification as in the top panel.}
\label{damping vs chi_1}   
\end{figure}

As a first illustration, we focus on the damping term $\mathcal{D}_{1,1}$ as defined in \eqref{damping term} with $l=|m|=1$. Using the values of the physical quantities as given in \tablename{}~\ref{table1} and a radius such as $r/r_t =0.05$, that is near the hydrogen-burning shell, it is plotted as a function of the normalized frequency $\sigma$ and $\chi_{1}$ in \figurename{}~\ref{damping vs chi_1}. For indication purposes, the frequency wave spectrum given in \eqref{spectrum func} is represented in the same figure with \smash{$B=\nu_p/N_t \sim 0.001$}, as expected from \tablename{}~\ref{table1}.

Figure~\ref{damping vs chi_1} shows that the wave damping is efficient for only two ranges of frequencies. Indeed, $\mathcal{D}_{1,1}$ is composed, on the one hand, of a negative extremum associated with retrograde IGW ($m=-1$), and in the other hand, of a positive one associated with prograde IGW ($m=1$) located at higher frequencies. The locations and amplitudes of both extrema depend on $\chi_{1}$. Whatever $\chi_1$, they are positioned on both sides of a characteristic frequency, denoted $\sigma_1^{{\rm sym}}$. It corresponds to the frequency at the extremum of both retrograde and prograde damping terms defined in \eqref{damping term 2} when $\Delta \Omega$ tends to zero (i.e., when retrograde and prograde IGW are symmetrically damped; see \eqref{omega sym} for a definition). 

We note that the effects of a change in the angular degree $l$ or in the value of $A_{l=1}$ can be easily deduced from $\mathcal{D}_{1,1}$ as displayed in \figurename{}~\ref{damping vs chi_1}. Indeed, following Eqs. (\ref{damping term}), (\ref{damping term 2}) and (\ref{tau 3}), a map of the $\mathcal{D}_{l,|m|}$ function in the ($\sigma$, $\chi_{|m|}$) plane associated with a parameter $A_{l}^\prime$ is deduced from the one associated with the parameter $A_l$, such as $A_l<A_l^{\prime}$, through a horizontal dilatation of the $\sigma$-axis by a factor \smash{$\alpha=(A_l^\prime/A_l)^{1/4}$} and a vertical contraction of the $\chi_{|m|}$-axis by a factor $1/\alpha$ (i.e.,~ensuring \smash{$A_l \chi_{|m|}^4 = \mathrm{cst}$}). In logarithmic scales, it corresponds to a translation $[0.25 \log\alpha,-0.25 \log\alpha]$ (see \figurename{}~\ref{damping vs chi_m 2} for an illustration). This explains our choice to first consider $l=|m|=1$ as example and to represent the effect of the variation in $\chi_{1}$ only .

In the following, we use the approximate expressions given in \sectionname{}~\ref{approximate I} to discuss the dependence of the damping term $\mathcal{D}_{l,|m|}$ on both parameters $A_l$ and $\chi_{|m|}$. The case $l=1$ displayed in \figurename{}~\ref{damping vs chi_1} is used as an illustration. Retrograde and prograde IGW are treated separately.

\subsubsection{Retrograde IGW $(m<0)$}
\label{retrograde analysis}

Near the hydrogen-burning shell, $z=r/r_t \ll 1$ and the retrograde part of \eqref{damping term} can thus be estimated with the help of \eqref{retrograde approximate}. Recalling that $|X|=\omega/|m|\Delta \Omega$ in \eqref{retrograde approximate}, the transition frequency between the two regimes given in \eqref{retrograde approximate} can be rewritten
\begin{align}
{\sigma_{t}^R}\sim\left( \dfrac{A_l \chi_{|m|} /3}{A_l \chi_{|m|} ^4 \left(r_t/r\right)^{\beta-1}/(\beta -1)}\right)^{1/3} \mbox{ .}
\label{transition retro}
\end{align}
Therefore, this makes us to consider the two following cases.

\paragraph{Case 1: Wave frequencies such as $\sigma>{\sigma_{t}^R}$.}
As discussed in \sectionname{}~\ref{outcome damping}, this range of frequencies corresponds to the wave frequencies for which the wave damping integral is dominated by the local damping.
Using \eqref{retrograde approximate} with $\sigma>\sigma_t^R$ in \eqref{tau 3}, the wave radiative damping can be approximated by
\begin{align}
\tau \approx \frac{A_l \chi_{|m|}^4}{(\beta -1)}\frac{\left(r_t/r\right)^{\beta-1} }{(\sigma \chi_{|m|}+1)^4}\mbox{  ,} \label{tau +}
\end{align}
and \eqref{damping term 2} for retrograde IGW ($m<0$) can be rewritten
\begin{align}
\widetilde{\mathcal{D}}_{l,m}\approx (\beta -1) \tau e^{-\tau} \mbox{  ,}\label{damping x sup}
\end{align}
where we have neglected $r/r_t$ compared to $\sigma \chi_{|m|}+1$ in the denominator.
Therefore, \smash{\smash{$\widetilde{\mathcal{D}}_{l,m}$}} is minimum for $\tau \approx 1$. Its peak amplitude is thus equal to \smash{$(\beta -1)/e$}. Following \eqref{tau +}, the frequency of retrograde IGW at maximum damping satisfying $\tau\approx1$ is given by
\begin{align}
\sigma_{{\rm max}}^R\approx  \sigma_l^{{\rm sym}} - \frac{1}{\chi_{|m|}}\mbox{ ,} \label{x_max retro}
\end{align}
provided that \smash{$\sigma_{{\rm max}}^R > {\sigma_{t}^R}$} and where we have defined
\begin{align}
 \sigma_l^{{\rm sym}} = \left(\frac{A_l }{\beta-1}\left(\frac{r_t}{r}\right)^{\beta-1} \right)^{1/4} \mbox{  .}\label{sigma sym}
\end{align}
This latter frequency corresponds to the frequency satisfying $\tau =1$ when $\chi_{|m|}\rightarrow +\infty$ (i.e., when the prograde and retrograde IGW suffered a symmetrical damping). We note that we can show that the full width at half maximum of the retrograde peak is about \smash{$0.6~\sigma_l^{{\rm sym}}$ around $\sigma_{{\rm max}}^R$}. Equations~(\ref{x_max retro}) and (\ref{sigma sym}) show that an increase either in $A_l$ (i.e., in the magnitude of the radiative diffusion) or in $r_t/r$ (i.e., in the distance covered by the waves) results in a shift of \smash{$\sigma_l^{{\rm sym}}$} and $\sigma_{{\rm max}}^R$ toward higher values. Equation~(\ref{x_max retro}) also tells us that a decrease in $\chi_{|m|}$ (i.e., an increase in the Doppler-shift) results in a shift of the retrograde damping peak toward lower frequencies. 
More precisely, \eqref{x_max retro} makes us to distinguish three qualitative regimes:
\begin{enumerate}
\item {\bf Case $\sigma_l^{{\rm sym}} \gg 1/\chi_{|m|} $ :} the retrograde damping peak is closely located near \smash{$\sigma_l^{{\rm sym}}$}, or in other words, $\sigma_{{\rm max}}^R\sim \sigma_l^{{\rm sym}}$. Using the values in \tablename{}~\ref{table1} and $l=1$, we found \smash{$\omega_{l=1}^{{\rm sym}}/N_t\sim0.034$}, in agreement with \figurename{}~\ref{damping vs chi_1}. 
\item {\bf Case $\sigma_l^{{\rm sym}}\sim 1/ \chi_{|m|} $ :} $\sigma_{{\rm max}}^R$ migrates toward lower frequencies as $\chi_{|m|}$ decreases (or $|m|\Delta \Omega$ increases).
\item {\bf Case $\sigma_l^{{\rm sym}}\ll 1/\chi_{|m|}$ :} $\sigma_{{\rm max}}^R$ becomes lower than $\sigma_t^R$. The wave frequencies such as $\sigma \gtrsim \sigma_t^R$ have not suffered an efficient damping yet since \smash{$\tau \approx (\sigma_l^{{\rm sym}} \chi_{|m|})^4 \ll 1$} from \eqref{tau +}. Therefore, following \eqref{damping x sup}, the retrograde damping term \smash{\smash{$\widetilde{\mathcal{D}}_{l,m}$}} rapidly vanishes as $\chi_{|m|}$ decreases.
\end{enumerate}
These three regimes are well observed in \figurename{} \ref{damping vs chi_1}.

\paragraph{Case 2: Wave frequencies such as $\sigma<\sigma_{t}^R$.}
This range of frequencies corresponds to the wave frequencies for which the wave damping integral is dominated by the contribution of the top of the radiative zone. Using \eqref{retrograde approximate} with $\sigma<\sigma_t^R$ in \eqref{tau 3}, we obtain
\begin{align}
\tau \approx \left( \frac{\sigma_{{\rm top}}{}}{\sigma}\right)^3 \mbox{  ,}
\label{tau + 2}
\end{align}
with
\begin{align}
\sigma_{{\rm top}}{}\approx \left(\frac{A_l\chi_{|m|}}{3}\right)^{1/3}=\left( \frac{[l(l+1)]^{3/2} K_t}{3 r_t^3}\frac{r_t}{ m \Delta\Omega}\right)^{1/3} \mbox{  .}
\label{sigma inf}
\end{align}
This latter is representative of the magnitude of the wave damping at the top of the radiative zone and of the competing effect between the radiative diffusion and the Doppler-shift in this region.
The larger the radiative diffusion (i.e., $K_t$) at the top of the radiative zone, the larger \smash{$\sigma_{{\rm top}}$} and $\tau$. Conversely, the larger the gradient in the wave intrinsic frequency (i.e., $m\Delta \Omega /r_t$), the lower \smash{$\sigma_{{\rm top}}$} and $\tau$ since $\hat{\omega}$ increases more rapidly as retrograde IGW propagate downwards.

If $\sigma< \sigma_t^R$, we can assume $\sigma \chi_{|m|}< \sigma_t^R \chi_{|m|}\ll 1$ using \eqref{transition retro} with $r/r_t\ll 1$. Then, by approximating the denominator of \eqref{damping term 2} by unity, we obtain
\begin{align}
\widetilde{\mathcal{D}}_{l,m}&\approx A_l \chi_{|m|}^4 \left(\dfrac{r_t}{r}\right)^{\beta-1} 
\exp\left(- \frac{\sigma_{{\rm top}}^3}{\sigma^3}\right) ~~~~\mbox{  .}\label{damping x inf}
\end{align}
Therefore, using Eqs. (\ref{transition retro}), (\ref{x_max retro}), (\ref{sigma sym}) and (\ref{sigma inf}), we can notice that
\begin{align}
\sigma_t^R=\frac{\sigma_{{\rm top}}{} }{(\chi_{|m|}\sigma_l^{{\rm sym}})^{4/3}}=\frac{\sigma_{{\rm top}}{} }{(\chi_{|m|}\sigma_{{\rm max}}^R+1)^{4/3}}\mbox{  ,}\label{relation sigma}
\end{align}
and we find using again \eqref{sigma sym} that \eqref{damping x inf} is approximately equal to
\begin{align}
\widetilde{\mathcal{D}}_{l,m}(\sigma< {\sigma_t^R},r)&\approx(\beta -1) \left(\frac{\sigma_{{\rm top}}{}}{\sigma_t^R}\right)^3\exp\left(- \frac{\sigma_{{\rm top}}^3}{\sigma^3}\right) \nonumber\\
&\lesssim (\beta -1) \left(\frac{\sigma_{{\rm top}}{}}{\sigma}\right)^3\exp\left(- \frac{\sigma_{{\rm top}}^3}{\sigma^3}\right) \mbox{  .}\label{damping x inf b}
\end{align}
Equation~(\ref{damping x inf b}) is thus a decreasing function of $\sigma$ that drastically vanishes as $\sigma$ becomes smaller than $\sigma_{{\rm top}}$. 
Indeed, following \eqref{tau + 2}, the wave damping integral is much higher than unity for frequencies such that $\sigma \lesssim \sigma_{{\rm top}}$ (i.e., $\tau \gg 1$). Thus, this range of low frequencies have already been significantly absorbed into the layers at the top of the radiative zone. We note that if \smash{$\sigma_l^{{\rm sym}}\ge 1/ \chi_{|m|} $}, or equivalently \smash{$\sigma_{{\rm max}}^R\ge0$}, \eqref{relation sigma} shows that \smash{$\sigma_t^R\le \sigma_{{\rm top}}{}$}. As a result, \eqref{damping x inf b} tends rapidly to zero as soon as $\sigma\le {\sigma_t^R}$. Otherwise, in the case \smash{$\sigma_l^{{\rm sym}}\le 1/ \chi_{|m|} $}, or equivalently \smash{$\sigma_{{\rm max}}^R\le0$}, retrograde IGW such as \smash{$\sigma_{{\rm top}}{}\le\sigma \le \sigma_t^R$} have not suffered an efficient damping yet since they verify $\tau \le 1$. We note also that if \smash{$\sigma_l^{{\rm sym}}\ge 1/ \chi_{|m|} $}, we always have \smash{$\sigma_l^{{\rm sym}}\ge \sigma_t^R$} following \eqref{transition retro} with $r/r_t\ll1$.
With the parameters of subgiant model given in \tablename{}~\ref{table1}, we find \smash{$\sigma_{{\rm top}} \approx 8~10^{-4}$} and \smash{$\sigma_t^R \approx 7~10^{-4}$ for $\chi_{|m|}=30$}, in agreement with \figurename{}~\ref{damping vs chi_1}. \\

{\it In summary, retrograde IGW deposit momentum into the deep layers of the radiative zone where their damping acoustic depth satisfy $\tau \approx1$ provided that \smash{$\sigma\gtrsim\sigma_t^R$}. This also implies that \smash{$\sigma_l^{{\rm sym}}\chi_{|m|} \gtrsim 1$}. IGW such as $\sigma_t^R\gtrsim\sigma$ or \smash{$\sigma_l^{{\rm sym}}\chi_{|m|} \ll 1$} either have already been absorbed at the top of the radiative zone or are not damped enough to locally deposit momentum.}

\subsubsection{Prograde IGW $(m>0)$}
\label{prograde analysis}
The prograde part of \eqref{damping term}, \smash{$\widetilde{\mathcal{D}}_{l,m}$}, can be estimated using \eqref{prograde approximate}. In the case $r/r_t\ll1$, the transition frequency between the two regimes for prograde waves, $\sigma_t^P$, can be defined by
\begin{align}
\sigma_t^P\chi_{|m|} -1 +r/r_t \sim \frac{3}{(\beta-1)} \frac{r}{r_t} \mbox{  .}
\end{align}
Then, injecting \eqref{prograde approximate} into \eqref{tau 3}, we obtain for $\sigma>\sigma_t^P$,
\begin{align}
\tau\approx \frac{A_l \chi_{|m|}^4}{\beta -1}\frac{\left(r_t/r\right)^{\beta-1} }{(\sigma \chi_{|m|}-1+r/r_t)^4}\mbox{  ,}\label{tau - 1}
\end{align}
and for $\sigma<\sigma_t^P$,
\begin{align}
\tau\approx \frac{A_l \chi_{|m|}^4}{3}\frac{\left(r_t/r\right)^{\beta} }{(\sigma \chi_{|m|}-1+r/r_t)^3} \mbox{  ,}\label{tau - 2}
\end{align}
on the condition that $\sigma \chi_{|m|} > 1-r/r_t$ (which is imposed by the presence of critical layers).
For our purpose, it is convenient to remark that $\sigma_t^P \sim 1/\chi_{|m|}$ with $r/r_t\ll1$, so that the transition frequency is very close to the frequency of the waves at their critical layers. In this framework, it is sufficient to consider only the case $\sigma_t^P \lesssim \sigma$ and the expression \eqref{tau - 1}. In this case, \eqref{damping term 2} for prograde IGW ($m>0$) can be rewritten such as
\begin{align}
\widetilde{\mathcal{D}}_{l,m}&\approx (\beta -1)~\tau e^{-\tau}\mbox{  .}\label{damping x sup pro}
\end{align}
Therefore, we conclude that the prograde part of the damping, \smash{$\widetilde{\mathcal{D}}_{l,m}$}, is maximum when $\tau \approx 1$ and that its peak amplitude is equal to $(\beta -1)/e$. The frequency at the maximum is thus equal to
\begin{align}
\sigma_{{\rm max}}^P\approx \sigma_{l}^{{\rm sym}} +\frac{(1-r/r_t)}{\chi_{|m|}}\mbox{  ,} \label{x_max pro large x_t}
\end{align}
with \smash{$ \sigma_{l}^{{\rm sym}} $} given by \eqref{sigma sym}.
Equation~(\ref{x_max pro large x_t}) shows that an increase in $A_l$ or $r_t/r$ leads to an increase in the frequency at the maximum. Similarly, the prograde peak migrates toward higher frequencies as $\chi_{|m|}$ decreases (i.e., $m \Delta \Omega$ increases), unlike the retrograde peak according to \eqref{x_max retro}. Moreover, for \smash{$\chi_{|m|}\sigma_{l}^{{\rm sym}} \gg 1$}, Eqs. (\ref{x_max retro}) and (\ref{x_max pro large x_t}) show that the prograde and retrograde peaks overlap around \smash{$ \sigma_{l}^{{\rm sym}} $}. Since they have identical values, both terms in \eqref{damping term} progressively cancel each other out as $\chi_{|m|}$ tends to infinity. Finally, for \smash{$\chi_{|m|}\sigma_{l}^{{\rm sym}} \ll 1$}, the prograde damping peak is associated with the waves that are dissipated close to their critical layers such as \smash{$\sigma_{{\rm max}}^P \sim \sigma_t^P\sim 1/\chi_{|m|}$}. All these features can be observed in \figurename{}~\ref{damping vs chi_1}.\\

{\it In summary, prograde IGW are absorbed into the deep layers of the radiative zone where they satisfy $\tau \approx 1$, that is either at frequencies around \smash{$\sigma_{l}^{{\rm sym}}$} if \smash{$\chi_{|m|}\sigma_{l}^{{\rm sym}} \gg1$} or at frequencies near their critical layers if \smash{$\chi_{|m|}\sigma_{l}^{{\rm sym}} \ll 1$}.}

\subsection{Key points for the transport by IGW near the hydrogen-burning shell}
\label{key points}

The analysis of the previous sections enables us to discuss in simple words the behavior of the transport by IGW in the vicinity of the hydrogen-burning shell. We summarize here the key points learned from the toy model. 

To begin with, we remind that the rotation profile is supposed to linearly decrease with respect to the radius and that the radius of the hydrogen-burning shell, $r_{{\rm HS}}$, is assumed to be much smaller than the radius at the top of the radiative zone, $r_t$, so that the differential rotation is nearly maximum in this region (i.e., $\delta \Omega(r\lesssim r_{{\rm HS}}) \approx \Delta \Omega$).

\subsubsection{Divergence of the wave flux.}
The angular momentum transport induced by IGW in \eqref{domega_dt} is carried out by the divergence of the wave flux of angular momentum, $\dot{J}_{{\rm w}}$. Following \eqref{J dot developed}, this latter results from the correlation between the wave excitation spectrum, represented by $\mathcal{S}$ and $\mathcal{F}_{l,|m|}$, and the wave radiative damping, represented by $\mathcal{D}_{l,|m|}$. Following Eqs.~(\ref{domega_dt})-(\ref{J_dot}), (\ref{wave energy flux})-(\ref{F_0}) and (\ref{total flux}) and the convention adopted in the footnote~\ref{foot 1} with $\delta \Omega(r) >0$, retrograde IGW ($m<0$) carrying negative angular momentum downwards tend to slow down the rotation since $\dot{J}_{{\rm w}}(m<0)>0$, while prograde IGW ($m>0$) carrying positive angular momentum downwards tend to increase the rotation rate since $\dot{J}_{{\rm w}}(m>0)<0$. 
\subsubsection{Damping efficiency and absorption criterion.}
To modify the rotation in a mass shell of radius $r\ll r_t$, IGW must be sufficiently damped to locally deposit angular momentum into the medium while having conserved enough energy during its travel from the top of the radiative zone to be efficient. Following Eqs. (\ref{x_max retro}) and (\ref{x_max pro large x_t}), the damping is efficient at a depth $r\ll r_t$ for the wave components whose the damping acoustic depth is close to unity
\begin{align}
\tau \approx \int_r^{r_t} \frac{[l(l+1)]^{3/2} K N^3}{(\omega-m \Delta \Omega)^4 } \frac{{\rm d }r}{r^3} \sim 1\mbox{  ,}
\label{damping unity}
\end{align}
with $-l \le m \le +l$. At fixed $l$ and $|m|\Delta \Omega$, lower frequencies such as $\tau \gg 1$ have already been partly absorbed into upper layers whereas higher frequencies such as $\tau \ll 1$ have not suffered an efficient damping yet during their travel. Moreover, since the rotation profile is assumed to decrease with the radius ($\Delta \Omega >0$), the prograde IGW ($m>0$) that are efficiently absorbed have always higher frequencies than the retrograde ones ($m<0$). For a given value of $\Delta \Omega$, since retrograde (prograde) IGW carry negative (positive) angular momentum, the damping term, $\mathcal{D}_{l,|m|}$ is therefore composed of a negative extremum associated with retrograde IGW and a positive one associated with prograde IGW located at a higher frequency. This is illustrated in \figurename{}~\ref{damping vs chi_1} for $l=1$ and $|m|=1$.

{\it To summarize, the prograde and retrograde IGW that are efficiently absorbed near the hydrogen burning shell satisfy $\tau \sim 1$.}

\subsubsection{Low cut-off frequency for retrograde IGW.}

For retrograde IGW, the condition of absorption represented by \eqref{damping unity} applies if the wave frequency is higher than a cut-off frequency, denoted $\omega_m^{{\rm inf}} $ in the following. In the framework of the toy model and \eqref{transition retro}, it is found to be equal to
\begin{align}
\omega_m^{{\rm inf}}(r) \approx \alpha(r)  |m|\Delta \Omega~~~~\mbox{with}~~~~\alpha(r)=\left[ \frac{(\beta -1)}{3 }\left(\frac{r}{r_t}\right)^{\beta-1}\right]^{1/3}
\mbox{ .}\label{omega t +}
\end{align}
It is defined as the wave frequency below which $\tau$ is dominated by the damping near the top of the radiative zone, and above which $\tau$ is dominated by the local damping. Therefore, if $\omega\lesssim\omega_m^{{\rm inf}} $, either IGW have already been absorbed by the medium at the top of the radiative zone, or they are not damped enough to efficiently deposit momentum into the local medium. Therefore, the damping function $\mathcal{D}_{l,|m|}$ drastically vanishes for the frequencies lower than \smash{$\omega_m^{{\rm inf}} $} (see \figurename{}~\ref{damping vs chi_1} for an illustration).

{\it Retrograde IGW meeting the condition $\tau \sim 1$ must also satisfy $\omega \gtrsim \omega_m^{{\rm inf}} $ to be efficiently absorbed near the hydrogen-burning shell. This defines the conditions of absorption for retrograde IGW. Conversely, prograde IGW are efficiently damped provided only that $\tau\sim 1$.}

\subsubsection{Effect of the frequency Doppler-shift on the wave damping.}
\label{effect of the Doppler-shift}

To describe the role of the differential rotation amplitude in the wave damping process, it is convenient to define the characteristic frequency, $\omega_{l}^{{\rm sym}}$, for which $\tau = 1$ is satisfied with $|m|\Delta \Omega=0$, that is
\begin{align}
\omega_{l}^{{\rm sym}}(r) \equiv \left( \int_r^{r_t} [l(l+1)]^{3/2} K N^3\frac{{\rm d }r}{r^3}\right)^{1/4}\mbox{  .}
\label{omega sym}
\end{align}
The condition $\tau \sim 1$ can then be rewritten as \smash{$\omega_{l}^{{\rm sym}}\sim(\omega-m\Delta \Omega)$}, which makes us to consider four cases (see \figurename{}~\ref{damping vs chi_1} for an example):
\begin{enumerate}
\item {\bf Case $\Delta \Omega = 0$:} In this limiting case, both prograde and retrograde damping peaks (i.e., where $\tau \sim 1$) of the $\mathcal{D}_{l,|m|}$ function overlap at \smash{$\omega =\omega_{l}^{{\rm sym}}$}. Since retrograde and prograde IGW are supposed to be symmetrically generated, they transport an opposite amount of angular momentum. Hence, the components cancel each other out such as $\mathcal{D}_{l,|m|}=0$ whatever $l$ and $|m|$; therefore, $\dot{J}_{{\rm w}}=0$ and the angular momentum transport by IGW vanishes, as expected (see also discussion in \sectionname{}~\ref{plume flux}).

\item {\bf Case $|m|\Delta \Omega \ll \omega_{l}^{{\rm sym}}$:} As $\Delta \Omega$ slightly increases, both peaks of the $\mathcal{D}_{l,|m|}$ function gradually separate but remain closely located on both sides of \smash{$\omega_{l}^{{\rm sym}}$} in this regime.

\item {\bf Case $|m|\Delta \Omega \gg \omega_{l}^{{\rm sym}}$:} In this regime, the Doppler-shift is so important that the wave damping is not efficient and \eqref{damping unity} cannot be satisfied for retrograde IGW (\smash{$\tau\ll 1$}). Only prograde IGW near their critical layers, such as \smash{$\omega\sim|m|\Delta \Omega$}, are locally absorbed.

\item {\bf Case $|m|\Delta \Omega \sim \omega_{l}^{{\rm sym}}$:} When $|m|\Delta \Omega$ is of the order of \smash{$\omega_l^{{\rm sym}}$} and gradually increases, the asymmetry between the prograde and retrograde damping peaks becomes more and more pronounced, as illustrated in \figurename{}~\ref{damping vs chi_1}. Indeed, the retrograde IGW satisfying $\tau \approx1$ at given $|m|$ and $l$ are shifted toward lower frequencies than \smash{$\omega_{l}^{{\rm sym}}$}, while the prograde ones migrate toward higher frequencies. In other words, at fixed $\omega$, $l$ and $|m|$, an increase in $\Delta \Omega$ leads to a decrease in the damping of retrograde IGW; they are thus absorbed into deeper layers of the star. In this regime, retrograde IGW are absorbed in the frequency range $\omega_m^{{\rm inf}}\lesssim\omega\lesssim\omega_l^{{\rm sym}} $. We note that for \smash{$|m|\Delta \Omega \lesssim \omega_{l}^{{\rm sym}}$}, we always have \smash{$\omega_m^{{\rm inf}}   \ll \omega_{l}^{{\rm sym}}$} with $r\ll r_t$ in \eqref{omega t +}, so that both conditions $\tau\approx1$ and \smash{$\omega_m^{{\rm inf}}\lesssim\omega\lesssim\omega_l^{{\rm sym}} $} can always be simultaneously satisfied in this regime. Conversely, an increase in $\Delta \Omega$ results in an increase in the damping of prograde IGW; they are thus absorbed into upper layers at frequencies \smash{$\omega\gtrsim\omega_l^{{\rm sym}} $}.

\end{enumerate}

{\it The condition of absorption $\tau \sim 1$ for retrograde IGW implies \smash{$|m|\Delta \Omega \lesssim \omega_{l}^{{\rm sym}}$}. When \smash{$|m|\Delta \Omega \gg \omega_{l}^{{\rm sym}}$}, only prograde IGW near their critical layers are absorbed. An increase in $\Delta \Omega$ leads to a shift of the retrograde (prograde) IGW that are absorbed toward lower (higher) frequencies; the asymmetry between prograde and retrograde IGW is thus accentuated.}

\subsubsection{Transport by IGW resulting from a balance between damping and driving.}
\label{damping vs driving}

In the following, we will assume that \smash{$\nu_p \ll \omega_{l=1}^{{\rm sym}}\le  \omega_{l}^{{\rm sym}}$} whatever $l$. Since $\nu_p$ is chosen around $\omega_c$, this is verified in almost all the models considered in the main text, except maybe at the very beginning of the subgiant branch (see Figs.~\ref{HR}~and~\ref{caract}).

\paragraph{Sign of $\dot{J}_{{\rm w}}$.}

Following \eqref{J dot developed}, the angular momentum transport by IGW and its efficiency depends on the correlation between the wave excitation spectrum and the wave damping. 
Excitation models show that the wave spectrum at the top of the radiative zone, $\mathcal{S}$ in \eqref{J dot developed}, is decreasing with the wave frequency, whatever the generation mechanism (turbulent pressure or penetrative convection). When \smash{$|m|\Delta \Omega \lesssim \omega_{l}^{{\rm sym}}$} at given $l$ and $|m|$, the retrograde IGW that are absorbed into a given layer of the helium core have lower frequencies than the prograde ones whose frequencies are always higher than \smash{$\omega_{l}^{{\rm sym}}$}. Given that $\mathcal{S}$ is a decreasing function of $\omega$, they have higher amplitudes compared to prograde IGW. Since they carry negative angular momentum, the total deposit of angular momentum is negative. Moreover, even if \smash{$|m|\Delta \Omega \ll \omega_{l}^{{\rm sym}}$} and both opposite absorption peaks are in the vicinity of \smash{$\omega_{l}^{{\rm sym}}$}, the amplitude of the retrograde IGW that are absorbed remains much higher than the amplitude of the prograde ones. Indeed, the full width at half maximum of the absorption peaks is about \smash{$0.6 ~\omega_{l}^{{\rm sym}}$} (see \appendixname{}~\ref{retrograde analysis}). Therefore, the separation in frequency between both prograde and retrograde peaks is large enough for the difference of amplitudes to remain important since the $\mathcal{S}$ function sharply decreases in this range of frequency with \smash{$\nu_p \ll\omega_{l}^{{\rm sym}}$}. Finally, when \smash{$|m|\Delta \Omega \gg \omega_{l}^{{\rm sym}}$}, only the prograde IGW near their critical layers are absorbed into the medium. However, since these waves are such that $\omega\sim |m|\Delta \Omega\ggg \nu_p$ in this regime, the $\mathcal{S}$ function drastically vanishes and their amplitude is quasi null, and so does their transport efficiency.

{\it The transport of angular momentum near the hydrogen-burning shell is ensured only by retrograde wave components. Therefore, the net wave flux of angular momentum directed downwards is negative, $\dot{J}_{{\rm w}}\gtrsim 0$ in \eqref{domega_dt}
and retrograde IGW tend to counter the contraction-driven spin up of the helium core, as expected.}

\paragraph{Influence of $\Delta \Omega$ on the transport by one wave component.}

At given $l$ and $|m|$, the higher $\Delta \Omega$, the lower the frequencies of the retrograde IGW that efficiently deposit momentum provided that \smash{$\omega_{l}^{{\rm sym}} \gtrsim |m|\Delta \Omega$}, and the higher the frequencies of the prograde ones. Since the excitation wave spectrum is a decreasing function of frequency, an increase in $\Delta \Omega$ leads to an increase in the amount of negative angular momentum absorbed into the medium. The effect on the rotation is thus more important. 

{\it At given $l$ and $|m|$, the efficiency of the transport by retrograde IGW increases as the amplitude of the Doppler-shift increases while remaining lower than about \smash{$\omega_{l}^{{\rm sym}}$}.} 

\paragraph{Influence of $\Delta \Omega$ on the whole spectrum.}
The previous scenario holds true if $|m|\Delta \Omega$ remains lower than \smash{$\omega_{l}^{{\rm sym}}$}. Actually, to properly understand the influence of the differential rotation amplitude on the transport by IGW, we need to take the collective effect of the whole wave angular degrees $l$ and azimuthal numbers such as $1\le|m|\le l$ into account, as well as their energy distribution represented by $\mathcal{F}_{l,|m|}$. We remind that the angular degree at the maximum of the \smash{$\mathcal{F}_{l,|m|}$} function can be estimated by \smash{$l_{{\rm max}} \sim r_t/b$}. Following \tablename{}~\ref{table1}, $l_{{\rm max}} \sim 20$ in the subgiant model used as example in this appendix. 
As an illustration, it is instructive to represent the sum of the damping terms over a few distinct values of $l$. Here, we consider only four angular degrees $l=1,25,100$ and $400$. The result is displayed in \figurename{}~\ref{damping vs chi_m 2} as a function of $\sigma$ and $\chi_{|m|}$. As already discussed in \appendixname{}~\ref{plot damping}, we retrieve that an increase in $l$ results in a shift of the related damping pattern toward lower $\chi_{|m|}$ values and higher frequencies. In the following discussion, we assume that \smash{$\nu_p\ll \omega_{l=1}^{{\rm sym}}$} and that the deposit of angular momentum is ensured only by retrograde IGW satisfying the absorption conditions, such as $\tau \sim 1$ (which implies \smash{$|m|\Delta \Omega \lesssim \omega_{l}^{{\rm sym}}$}) and $\omega \gtrsim \omega_m^{{\rm inf}} $. As a result, following Eqs. (\ref{J dot developed}) and (\ref{damping term}), and using \eqref{damping x sup} with $\tau \sim 1$, the contribution of each of these components to the divergence of the wave flux is about proportional to the product $\mathcal{F}_{l,m} \mathcal{S}(\omega)$. Therefore, \figurename{}~\ref{damping vs chi_m 2} makes us to consider three regimes classified by ascending value of $\Delta \Omega$:
\begin{enumerate}

\item {\bf Very low-$\Delta \Omega$ regime:}

\vspace{0.2cm}
The case \smash{$|m|\Delta \Omega \sim \omega_{l}^{{\rm sym}}\sim l^{3/4}\omega_{l=1}^{{\rm sym}}$}, as described in \appendixname{}~\ref{effect of the Doppler-shift}, can be satisfied (with $|m|\le l$) by angular degrees and azimuthal numbers such as \smash{$l \ge |m| \gtrsim( \omega_{l=1}^{{\rm sym}}/\Delta \Omega)^4$}. This also implies $\omega_m^{{\rm inf}}\approx\alpha |m|\Delta \Omega \sim\alpha  l^{3/4}\omega_{l=1}^{{\rm sym}} $.
Therefore, in this first regime, we assume that the amplitude of the differential rotation is so low that \smash{$|m|\Delta \Omega \sim \omega_{l}^{{\rm sym}}$} is only satisfied for very high values of $l$ such as \smash{$l \gg l_{{\rm max}}$} and \smash{$\omega_m^{{\rm inf}} \gg \omega_{l=1}^{{\rm sym}}$} (see for example $l=400$ in \figurename{}~\ref{damping vs chi_m 2}), that is
\begin{align}
\Delta  \Omega \ll \omega_{l=1}^{{\rm sym}}  \min\left(l_{{\rm max}}^{-1/4},\alpha^{1/3}\right)\mbox{  .}
\end{align}
In turn, angular degrees such as \smash{$l \lesssim ( \omega_{l=1}^{{\rm sym}}/\Delta \Omega)^4$} can satisfy only the case \smash{$|m|\Delta \Omega \ll \omega_{l}^{{\rm sym}}$} with a peak of absorption around the frequency \smash{$\omega_{l}^{{\rm sym}}$}.

On the one hand, since \smash{$\mathcal{F}_{l,|m|}$} and $\mathcal{S}$ sharply decrease beyond $l_{{\rm max}}$ and $\nu_p$ respectively, the amplitude of the wave components satisfying \smash{$|m|\Delta \Omega \sim \omega_{l}^{{\rm sym}}$}, such as $l\gg l_{{\rm max}}$ and \smash{$\omega_m^{{\rm inf}}  \gtrsim \omega_{l=1}^{{\rm sym}}\gg \nu_p$} in this regime, as well as the conditions of absorption $\tau\sim 1$ and \smash{$\omega \gtrsim \omega_m^{{\rm inf}}$}, is negligible compared to the low degrees $l \sim 1$ satisfying \smash{$|m|\Delta \Omega \ll \omega_{l}^{{\rm sym}}$}, $\tau \sim 1$ and so \smash{$\omega \sim \omega_{l=1}^{{\rm sym}}$}.

On the other hand, among the other wave components satisfying \smash{$|m|\Delta \Omega \ll \omega_{l}^{{\rm sym}}$}, as well as $\tau\sim 1$ with \smash{$ \omega\sim  \omega_{l}^{{\rm sym}}$}, the deposit of angular momentum into the medium is again mainly ensured by retrograde IGW with low angular degrees $l \sim 1$ such as \smash{$\omega \sim \omega_{l=1}^{{\rm sym}}$} since the product \smash{$\mathcal{F}_{l,|m|}\mathcal{S}(  \omega_{l}^{{\rm sym}})$} sharply increases as $l$ decreases. Indeed, given that \smash{$\nu_p\ll \omega_{l=1}^{{\rm sym}}\le \omega_{l}^{{\rm sym}}$}, we can show for $l\lesssim l_{{\rm max}}$ that the decrease in $\mathcal{F}_{l,|m|}$ as $l$ decreases can be largely outweighed by the increase in $\mathcal{S}$ evaluated at \smash{$\omega \sim \omega_{l}^{{\rm sym}}$}.

{\it Therefore, low angular degrees $l\sim 1$ with frequencies around \smash{$\omega \sim \omega_{l=1}^{{\rm sym}}$} provide, in this regime, the main contribution to the total divergence of the wave flux, $\dot{J}_{{\rm w}}$.}
\\
\item  {\bf Moderate-$\Delta \Omega$ regime:}

\vspace{0.2cm}

\noindent As $\Delta \Omega$ increases, \smash{$|m|\Delta \Omega \sim \omega_{l}^{{\rm sym}}\sim l^{3/4}\omega_{l=1}^{{\rm sym}}$} with $|m| \le l$ can be satisfied by lower and lower azimuthal numbers at a given $l$, as well as by lower and lower angular degrees with lower and lower cut-off frequencies since \smash{$\omega_m^{{\rm inf}}\approx\alpha |m|\Delta \Omega \sim\alpha  l^{3/4}\omega_{l=1}^{{\rm sym}} $} in this case.

In a first step, we consider only the wave components such as $\omega \gtrsim \nu_p$ or $l\gtrsim l_{{\rm max}}$. Among these wave components satisfying \smash{$|m|\Delta \Omega \sim \omega_{l}^{{\rm sym}}$}, $\tau \sim 1$ and \smash{$ \omega_m^{{\rm inf}}\lesssim \omega $}, the lowest angular degrees with $|m|\sim l$ and \smash{$\omega \sim \omega_l^{{\rm inf}} $} have the highest amplitude. Indeed, on the one hand, $\mathcal{S}$ decreases with frequency, so that the waves with frequencies \smash{$\omega \sim \omega_m^{{\rm inf}}$} have the largest amplitude at given $l$ and $|m|$. On the other hand, the product \smash{$\mathcal{F}_{l,|m|} \mathcal{S}(\omega_m^{{\rm inf}})$} with \smash{$\omega_m^{{\rm inf}}\sim \alpha l^{3/4} \omega_{l=1}^{{\rm sym}} $} sharply increases as $l$ decreases if $\omega \gtrsim \nu_p$ or $l \gtrsim l_{{\rm max}}$, since the variations in the Gaussian terms of $\mathcal{F}_{l,|m|}$ or $\mathcal{S}$ predominate. Hence, the lowest degrees among the components satisfying \smash{$|m|\Delta \Omega \sim \omega_{l}^{{\rm sym}}$}, so with $|m| \sim l$, have the highest amplitude. Therefore, when $\Delta \Omega$ becomes large enough, the conditions \smash{$l\Delta \Omega \sim \omega_{l}^{{\rm sym}}$}, $\tau \sim 1$ and \smash{$\omega \sim \omega_l^{{\rm inf}} $} can be satisfied by angular degrees such as the product \smash{$\mathcal{F}_{l,l} \mathcal{S}(\omega_l^{{\rm inf}})\gtrsim\mathcal{F}_{1,1} \mathcal{S}(\omega_{l=1}^{{\rm sym}})$}, so that they have a higher amplitude than the components $l\sim 1$ with \smash{$\omega \sim \omega_{l=1}^{{\rm sym}}$}.These components thus start giving the most effective contribution to the angular momentum transport: this corresponds to the transition from the very low-$\Delta \Omega$ to the moderate-$\Delta \Omega$ regime. Using the inequality \smash{$\mathcal{F}_{l,l} \mathcal{S}(\omega_l^{{\rm inf}})\gtrsim\mathcal{F}_{1,1} \mathcal{S}(\omega_{l=1}^{{\rm sym}})$} with \smash{$\nu_p \ll \omega_{l=1}^{{\rm inf}}$}, we find that this regime corresponds to values of $\Delta \Omega$ such as\footnote{We also used the relations \smash{$\omega \sim \alpha l^{3/4} \omega_{l=1}^{{\rm sym}}$} and \smash{$l\sim (\omega_{l=1}^{{\rm sym}}/\Delta \Omega)^4$} (obtained with $|m|\sim l$).}
\begin{align}
\omega_{l=1}^{{\rm sym}} ~\max\left[\left(\frac{\nu_p}{\omega_{l=1}^{{\rm sym}} l_{{\rm max}}}\right)^{1/4},\alpha^{1/3}\right]\lesssim\Delta\Omega\lesssim \omega_{l=1}^{{\rm sym}}\mbox{  ,}\label{inequality 2}
\end{align}
with the upper limit obtained in the limiting case where the conditions \smash{$l\Delta \Omega \sim \omega_{l}^{{\rm sym}}$} is satisfied for $l=1$.
In this regime, with in addition the condition $\omega \gtrsim \nu_p$ or $l\gtrsim l_{{\rm max}}$, the transport of angular momentum is thus ensured by retrograde IGW such as 
\begin{align}
|m|\sim l~~~~,~~~~\omega\sim \omega_l^{{\rm inf}}\sim\alpha l^{3/4} \omega_{l=1}^{{\rm sym}}~~~~,~~~~l \sim \left( \frac{\omega_{l=1}^{{\rm sym}}}{\Delta \Omega}\right)^4 \mbox{  .}\nonumber
\end{align}
In this case, a gradual increase in $\Delta \Omega$ extends the domain of the ($l$,$\omega$) plan satisfying the conditions of absorption toward lower and lower angular degrees and wave frequencies.
Therefore, the sum in \eqref{J dot developed} as well as the total amount of angular momentum deposited into the medium increases since the product \smash{$\mathcal{F}_{l,l} \mathcal{S}(\omega_l^{{\rm inf}})$} increases as $l$ decreases in this case.

In a second step, we consider only the wave components such as $\omega \lesssim \nu_p$ and $l\lesssim l_{{\rm max}}$ (see for example $l=1$ in \figurename{}~\ref{damping vs chi_m 2}). These components can satisfy the case \smash{$|m|\Delta \Omega \sim \omega_{l}^{{\rm sym}}$} as soon as\footnote{We note that since \smash{$\nu_p\ll \omega_{l=1}^{{\rm sym}}$}, \eqref{inequality 3} is included in \eqref{inequality 2}.}
\begin{align}
\omega_{l=1}^{{\rm sym}} ~\max\left[ l_{{\rm max}}^{-1/4},\left(\frac{\omega_{l=1}^{{\rm sym}}  \alpha}{\nu_p}\right)^{1/3}\right]\lesssim\Delta\Omega
\mbox{  .}\label{inequality 3}
\end{align}
When $l\lesssim l_{{\rm max}}$ and $\omega \lesssim \nu_p$, the product \smash{$\mathcal{F}_{l,m} \mathcal{S}(\omega_m^{{\rm inf}})$} decreases as $l$ decreases since the decrease in the quadratic term of $\mathcal{F}_{l,|m|}$ predominates. Therefore, this time, the wave components with the highest $l$ satisfying $\tau\sim 1$, \smash{$|m|\Delta \Omega \sim \omega_{l}^{{\rm sym}}$} and \smash{$\omega \sim \omega_m^{{\rm inf}}\sim \alpha l^{3/4} \omega_{l=1}^{{\rm sym}}$}, with $|m|\le l$, are expected to keep the highest amplitude as $\Delta \Omega$ decreases. Under the condition $l\lesssim l_{{\rm max}}$ and $\omega \lesssim \nu_p$, the angular momentum transport is then ensured by wave components such as 
\begin{align}
\omega \sim \min\left[ \nu_p,\alpha l_{{\rm max}}^{3/4} \omega_{l=1}^{{\rm sym}}\right]~~~~\mbox{and}~~~~l\sim \min\left[(\nu_p/ \alpha \omega_{l=1}^{{\rm sym}})^{4/3}, l_{{\rm max}}\right] \nonumber
\end{align}
In this case, the amount of angular momentum deposited into the medium must remain almost constant as $\Delta \Omega$ decreases.

{\it In summary, the transition from the very low-$\Delta \Omega$ to the moderate-$\Delta \Omega$ regime is associated with an increase in the angular degrees of IGW that mostly contribute to the transport. Their angular degrees are such as 
\begin{align}
l\sim \max\left\lbrace \min\left[\left(\frac{\nu_p}{ \alpha \omega_{l=1}^{{\rm sym}}}\right)^{4/3}, l_{{\rm max}}\right],\left(\frac{\omega_{l=1}^{{\rm sym}}}{\Delta \Omega}\right)^4 \right\rbrace \mbox{  ,}
\end{align}
and their frequencies are such as 
\begin{align}
\omega  \sim \max\left\lbrace \min\left[\nu_p,  \alpha l_{{\rm max}}^{3/4} \omega_{l=1}^{{\rm sym}}\right], \alpha \omega_{l=1}^{{\rm sym}}\left(\frac{\omega_{l=1}^{{\rm sym}}}{\Delta \Omega}\right)^3  \right\rbrace\mbox{  .}
\end{align}
In this regime, an increase in $\Delta \Omega$ results in an increase in the transport efficiency through the contribution of wave components with lower and lower frequencies and angular degrees.}
\\ 
\item {\bf Very high-$\Delta \Omega$ regime:}

\vspace{0.2cm}

\noindent As $\Delta \Omega$ becomes such as 
\begin{align}
\omega_{l=1}^{{\rm sym}}\gg\Delta \Omega \mbox{  }\label{inequality 4}
\end{align}
and keeps on increasing, more and more wave components start satisfying the case \smash{$|m|\Delta \Omega \gg \omega_{l}^{{\rm sym}}$} and thus cannot be absorbed anymore into the medium. The conditions of absorption are thus satisfied in a smaller and smaller domain in the $(l,\omega)$ plan reduced to higher and higher degrees and higher and higher frequencies, with $|m|\lneqq l$ since \smash{$l\Delta \Omega \sim \omega_{l}^{{\rm sym}}$} has already been satisfied for a lower value of $\Delta \Omega$.

{\it Therefore, as $\Delta \Omega$ keeps on increasing and becomes such as \smash{$\Delta \Omega \gg \omega_{l=1}^{{\rm sym}}$}, $\dot{J}_{{\rm w}}$ and the efficiency of the transport by IGW gradually decreases. }
\end{enumerate}
The previous analysis is sufficient to qualitatively understand the influence of $\Delta \Omega$ on the efficiency of the angular momentum transport by IGW a well as the features of the most efficient wave components. A quantitative estimate taking the whole spectrum over the frequency, the angular degrees, as well as the sum over azimuthal numbers into account requires a numerical computation, as done in the main text. For all the considered models and values of the differential rotation amplitude, we found that only the low-$\Delta \Omega$ and moderate-$\Delta \Omega$ regimes were reached in this work and that their above qualitative description agrees with numerical results. In both regimes, an increase in $\Delta \Omega$ leads to an increase in the efficiency of the transport by IGW in the helium core, as observed in \sectionname{}~\ref{comparison}.
\begin{figure}
\centering
\includegraphics[scale=0.55,trim= 0.5cm 0cm 0cm 0cm, clip]{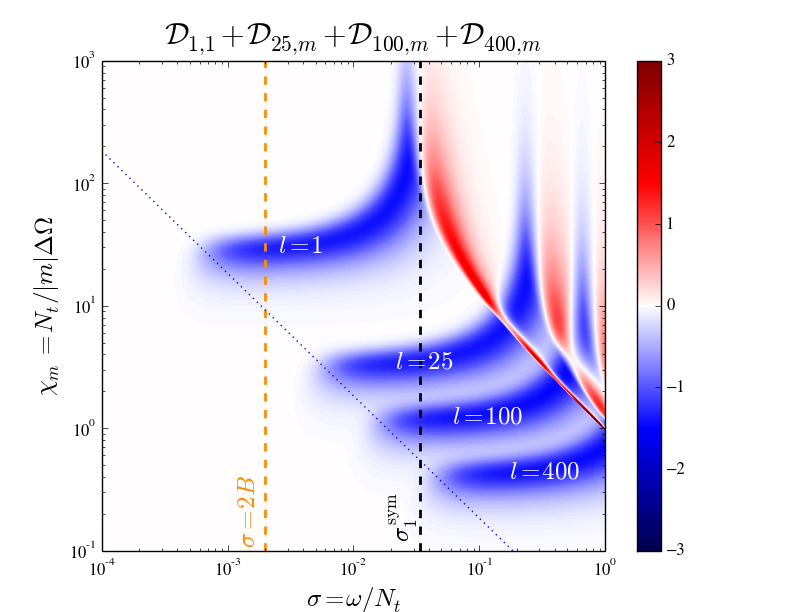}
\caption{Sum of the dimensionless damping terms defined by \eqref{damping term} over $l=1,25,100$ and $400$, as a function of $\chi_{|m|}$ and $\sigma=\omega/N_t$. The legend is the same as in the top panel of \figurename{}~\ref{damping vs chi_1}. The dotted line represents $\omega_m^{{\rm inf}}/N_t$ as defined in \eqref{omega t +}. The angular degree associated with each pattern is indicated.}
\label{damping vs chi_m 2}   
\end{figure}

\subsubsection{Absorbed wave components for a differential rotation amplitude close to the threshold.}
\label{absorbed component}

Using the value of the threshold amplitude for the differential rotation obtained with a numerical computation, we can crudely estimate the characteristics of the wave components that participate the most efficiently to the angular momentum transport into the helium core. As an example, we consider again the subgiant model whose parameters are provided by \tablename{}~\ref{table1}. For this model, computations presented in the main text give $\Delta \Omega_{{\rm th}}~\approx~5~\upmu$rad~s$^{-1}$. We supposed that this model is in the moderate-$\Delta \Omega$ regime (see \appendixname{}~\ref{damping vs driving}) and that the absorbed waves have frequencies such as $\omega \gtrsim \nu_p$. Therefore, we can first estimate the angular degrees of IGW that deposit angular momentum near the hydrogen burning shell. In this regime (with $\omega \gtrsim \nu_p$), the most efficiently absorbed wave components must satisfy $\tau\sim 1$, $|m|\sim l$ and \smash{$\omega \sim \omega_{l}^{{\rm inf}}$}. Given that $ \omega_{l}^{{\rm inf}}\ll \omega_{l}^{{\rm sym}}$ following the point 4 in \appendixname{}~\ref{effect of the Doppler-shift}, we must have $\omega\ll l\Delta \Omega$ to satisfy $\tau \sim 1$. Thus, we obtain using the condition $\tau \sim 1$
\begin{align}
l \sim \int_r^{r_t}  \frac{K N^3}{ \Delta \Omega_{{\rm th}}^4 } \frac{{\rm d }r}{r^3} \sim \frac{K_t N_t^3}{4 r_t^2\Delta \Omega_{{\rm th}}^4}\left( \frac{r_t}{r_{{\rm HS}}}\right)^4\sim 60 \mbox{  .}
\end{align}
Using this value, the low cut-off frequency defined by \eqref{omega t +} is then equal to about \smash{$\omega_l^{{\rm inf}} =6~\upmu$rad s$^{-1} \gg \nu_p$} and so is representative of the wave frequency $\omega$ that deposits the more angular momentum into the medium. To conclude, we verify a posteriori the validity of the assumptions of the moderate-$\Delta \Omega$ regime given in \eqref{inequality 2} and $\omega \gtrsim \nu_p$ for $\Delta \Omega= \Delta \Omega_{{\rm th}}$ since \smash{$\omega_{l=1}^{{\rm sym}}\sim 18~\upmu {\rm rad~s}^{-1}$} and \smash{$\alpha^{1/3}\sim \nu_p/l_{{\rm max}} \omega_{l=1}^{{\rm sym}} \sim 0.2$} (using \tablename{}~\ref{table2}). We checked that these values of $l$ and $\omega$ are in good agreement with the numerical computation for the considered subgiant model.

{\it This emphasizes that high-$l$, high-$|m|$ retrograde IGW with $\omega \gg \nu_p$ can be responsible for the deposit of angular momentum into the helium core of the low-mass stars on the subgiant and early red giant branch.}

\subsubsection{Influence of the damping magnitude.}
\label{influence of damping}
The magnitude of the damping is represented by the ratio \smash{$[l(l+1)]^{3/2}KN^3/r^3$} (see \appendixname{}~\ref{wave radiative damping}). If this latter increases at a given value of $|m|\Delta \Omega$, the wave frequencies satisfying the absorption conditions will increase for both retrograde and prograde IGW. The part of the wave spectrum that is susceptible to locally affect the rotation in a given mass shell will be thus shifted toward higher frequencies. Since the wave excitation spectrum is a decreasing function of $\omega$, the amplitude of the wave components absorbed into the medium and the transport efficiency ($\dot{J}_{{\rm w}}$) will decrease. 
Nevertheless, we have shown in \appendixname{}~\ref{effect of the Doppler-shift} that, for retrograde IGW at given $l$, $|m|$ and $\omega$, an increase in the magnitude of the damping can be counterbalanced by an increase in the amplitude of the differential rotation (provided also that \smash{$|m|\Delta \Omega \lesssim \omega_{l}^{{\rm sym}}$} and $\omega \gtrsim \omega_m^{{\rm inf}} $). Thus, for similar wave excitation spectra $\mathcal{F}_{l,|m|}$ and $\mathcal{S}$ in \eqref{J dot developed} (i.e., similar plume parameters), the higher the magnitude of the damping, the higher the value of $\Delta \Omega$ required to conserve the same value of $\dot{J}_{{\rm w}}$ and thus of $t_{{\rm w}}$ in \eqref{t_w} near the hydrogen-burning shell. For the 1~M$_\odot$ evolutionary sequence considered in this work, $t_{\rm cont}$ remains quasi constant around 0.5~Gyr and the radius of the hydrogen-burning shell only slightly changes from the subgiant to the early red giant branch (see \figurename{}~\ref{timescale sg rg}). Nevertheless, \figurename{}~\ref{caract} show that $\nu_p$ increases by about a factor of two and that $l_{{\rm max}}$ decreases by about a factor of ten. If we assume that the magnitude of the radiative damping does not change over time, the results obtained in \sectionname{}~\ref{uncertainties} show that such variations in $\nu_p$ and $l_{{\rm max}}$ cannot explain the increase in the value of $\Delta \Omega$ by about a factor of five that is required to conserve $t_{{\rm w}} \sim t_{{\rm cont}}$ in the helium core during this time span (see Figs.~\ref{threshold var}~and~\ref{loop}). Therefore, we conclude that an increase in the magnitude of the radiative damping is needed to explain such a trend. This increase is well observed in stellar models (see \appendixname{}~\ref{wave radiative damping}).

{\it The increase in the threshold value for the differential rotation amplitude over time, as observed in \figurename{}~\ref{loop}, mainly results from the increase in the magnitude of the radiative damping along the evolution.}

\subsubsection{Sensitivity of $\dot{J}_{{\rm w}}$ to the differential rotation amplitude.}
\label{sensitivity}
The high sensitivity of the wave-related timescale to $\Delta \Omega$ observed in \sectionname{}~\ref{IGW transport} can be explained by two main points. First, since the wave radiative damping depends on the wave intrinsic frequency to the power minus four, the full width at half maximum of the absorption peaks for given values of $l$ and $|m|$ is quite low so that they remain confined around the wave frequency satisfying $\tau =1$. Second, a small change in $\Delta \Omega$ can lead to a huge variation in the wave spectrum functions evaluated at the maximum of absorption. Indeed, guided by numerical computations and the analysis in \appendixname{}~\ref{absorbed component}, we can assume to be in the moderate-$\Delta \Omega$ regime and that the absorbed wave components are such as $\omega \gtrsim \nu_p$ or $l \gtrsim l_{{\rm max}}$. In this regime, IGW that the most efficiently participate to the transport in a given layer of radius $r$ satisfy $\tau \sim 1$, $|m|\sim l$ and \smash{$\omega\sim \omega_l^{{\rm inf}}\sim \alpha l \Delta \Omega $} (see \appendixname{}~\ref{damping vs driving}). This also implies \smash{$l\Delta \Omega \sim \omega_{l}^{{\rm sym}}$} or equivalently, \smash{$l^{1/4}\Delta \Omega \sim \omega_{l=1}^{{\rm sym}}$}. Hence, it is possible to express $\omega$ and $l$ of the absorbed wave components as a function of $\Delta \Omega$ only to obtain
\begin{align}
\omega &\sim \alpha \omega_{l=1}^{{\rm sym}} \left( \frac{ \omega_{l=1}^{{\rm sym}} }{\Delta \Omega}\right)^3\\
l&\sim \left( \frac{ \omega_{l=1}^{{\rm sym}} }{\Delta \Omega}\right)^4 \mbox{ .}
\end{align}
Therefore, the frequency $\omega$ and the degree $l$ of the most efficiently absorbed components are quite sensitive to the value of $\Delta \Omega$.
A small relative variation in $\Delta \Omega$ thus can lead to a huge variation in the $\mathcal{S}$ (or $\mathcal{F}_{l,|m|}$) function at the maximum of absorption if $\omega \gtrsim \nu_p$ (or $l\gtrsim l_{{\rm max}}$) since it depends on \smash{$\exp[-\omega^2/4\nu_p^2]$} (or \smash{$\exp[-l^2/2l_{{\rm max}}^2]$}). Equivalently, since the sensitivity of $t_{{\rm w}}$ is linked to $\mathcal{S}$ and $\mathcal{F}_{l,|m|}$, we deduce that a small variation in $\nu_p$ or $b$ at fixed $\Delta \Omega$ must result in a huge variation in $t_{{\rm w}}$. This is confirmed in \figurename{}~\ref{timescale var}. For this model, we observe that the effect of a variation in $\nu_p$ is much larger than the one induced by a variation in $b$. This is expected since $\omega \gg \nu_p$ and $l \sim l_{{\rm max}}$ for the most efficient wave components (see \appendixname{}~\ref{absorbed component}) so that the variation in $\mathcal{S}$ is larger.

{\it Therefore, a small increase in $\Delta \Omega$ can result in a huge increase in $\dot{J}_{{\rm w}}$ and so in the efficiency of the angular momentum transport by IGW in the helium core.}

\end{document}